\documentclass[sigconf]{acmart}

\usepackage[utf8]{inputenc} 
\usepackage[T1]{fontenc}    
\usepackage{amsfonts}       
\usepackage{nicefrac}       
\usepackage{microtype}      
\usepackage{xcolor}         %
\usepackage{siunitx}
\usepackage{subcaption}

\usepackage{enumitem}
\usepackage{multirow}
\usepackage{balance}
\usepackage{acmart-taps}

\newcommand{\xref}[1]{\S\ref{#1}}

\newcommand{\squishlist}{\begin{itemize}[itemsep=1pt,parsep=2pt,topsep=3pt,partopsep=0pt,leftmargin=0em, itemindent=1em,labelwidth=1em,labelsep=0.5em]}
\newcommand{\squishend}{\end{itemize}}

\newenvironment{todo-env}{\par\color{red}}{\par}
\newenvironment{help-env}{\par\color{blue}}{\par}
\newenvironment{ready-for-review}{\par\color{violet}}{\par}

\newif\ifcomments
\commentstrue 
\ifcomments
    \newcommand{\steve}[1]{\textcolor{cyan}{[SS: #1]}}
    \newcommand{\shyam}[1]{\textcolor{magenta}{[SG: #1]}}
    \newcommand{\ira}[1]{\textcolor{red}{[I: #1]}}
\else
    \providecommand{\steve}[1]{}
    \providecommand{\shyam}[1]{}
    \providecommand{\ira}[1]{}
\fi

\copyrightyear{2023}
\acmYear{2023}
\setcopyright{rightsretained}
\acmConference[UIST '23]{The 36th Annual ACM Symposium on User Interface Software and Technology}{October 29-November 1, 2023}{San Francisco, CA, USA}
\acmBooktitle{The 36th Annual ACM Symposium on User Interface Software and Technology (UIST '23), October 29-November 1, 2023, San Francisco, CA, USA}
\acmDOI{10.1145/3586183.3606779}
\acmISBN{979-8-4007-0132-0/23/10}

\begin{document}


















\title{{Semantic Hearing: Programming Acoustic Scenes  with  \\Binaural~Hearables }}
 







 \author{Bandhav Veluri}\authornote{Co-primary student authors}
  \affiliation{Paul G. Allen School, University of Washington, Seattle, WA  
 \country{USA}
 }
 \email{bandhav@cs.washington.edu}

 \author{Malek Itani}\authornotemark[1]
  \affiliation{Paul G. Allen School, University of Washington, Seattle, WA  
 \country{USA}
 }
 \email{malek@cs.washington.edu}

 \author{Justin Chan}
 \affiliation{Paul G. Allen School, University of Washington, Seattle, WA  
 \country{USA}
 }
 \email{jucha@cs.washington.edu }

 \author{Takuya Yoshioka}
 \affiliation{Microsoft, One Microsoft Way, Redmond, WA
  \country{USA}
 }
 \email{tayoshio@microsoft.com}

 \author{Shyamnath Gollakota}
 \affiliation{Paul G. Allen School, University of Washington, Seattle, WA
   \country{USA}
 }
\email{gshyam@cs.washington.edu}






 \renewcommand{\shortauthors}{Bandhav Veluri, Malek Itani, Justin Chan, Takuya Yoshioka and Shyamnath Gollakota}

 \renewcommand{\shortauthors}{Veluri, Itani, Chan,  Yoshioka and Gollakota}

\begin{abstract}
Imagine being able to listen to the birds chirping in a park without hearing the chatter from other hikers, or being able to block out traffic noise on a busy street while still being able to hear emergency sirens and car honks.    {We introduce {\it semantic hearing}, a novel capability for hearable devices that enables them to, in real-time, focus on, or ignore, specific sounds from real-world environments, while also preserving the spatial cues.} To achieve this, we make two technical contributions: 1)  we present the  first neural network that can achieve binaural target sound extraction in the presence of interfering sounds and background noise, and 2) we  design a training methodology that allows our system to generalize to real-world use. Results show that our system can operate with 20 sound classes and that our transformer-based network has a runtime of 6.56 ms on a connected smartphone. In-the-wild evaluation with participants in previously unseen indoor and outdoor scenarios shows  that our proof-of-concept system can extract the target  sounds and  generalize  to preserve the spatial cues in its binaural output. \vskip 0.05in \noindent {Project page with code: {\textcolor{blue}{{{\url{https://semantichearing.cs.washington.edu}}}}}}
\end{abstract}



\begin{teaserfigure}
    \centering
    \includegraphics[width=\textwidth]{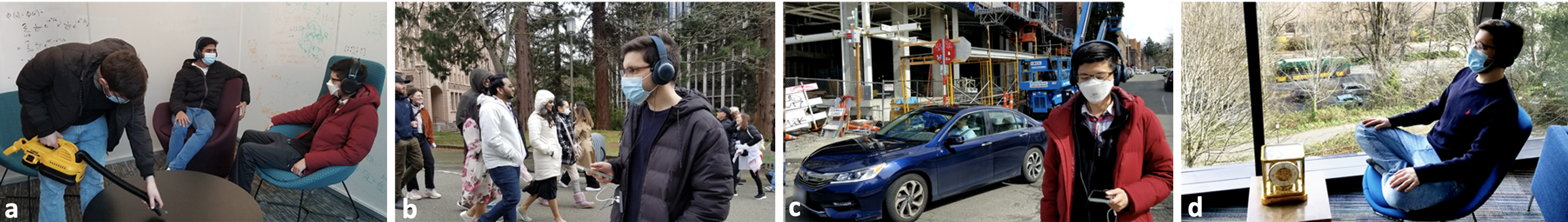}
    \caption[]{Semantic hearing applications.   a) Users wearing binaural headsets can attend to speech while blocking out only the vacuum cleaner noise, b) block out street chatter and focus on the sounds of birds chirping, c) block out construction noise yet hear car honks, and d) a meditating user could use  headsets to block out  traffic noise outside yet hear alarm clock sounds.}
    \label{fig:fig1}
    \vskip 0.1in
\end{teaserfigure}

\maketitle

\title{{Semantic Hearing: Programming  Acoustic Scenes  with  Binaural Hearables}}

\section{Introduction}




Over the past decade, we have witnessed an  increase in the number of hearable  devices like headsets, and  earbuds, with  millions of people  using them worldwide~\cite{airpodssales}.  Here, we  introduce  a new capability for hearable devices, which we call {\it``semantic hearing"}.


Consider a scenario where a user is wearing ear-worn devices  on a beach and desires to listen  to the calming sounds of the ocean while blocking out any human speech nearby. Similarly, while walking on a busy street, the user may wish to reduce all sounds except for emergency sirens; or while sleeping, they may want to listen to the alarm clock or baby sounds but not the noise from the street. In another scenario, the user may be on a plane and desire to hear human speech and announcements but not the sound of a crying baby. Or while hiking, the user may want to listen to the birds chirping but not the chatter from other hikers (see examples in Fig.~\ref{fig:fig1}). 
These and other potential use cases require noise-canceling earphones for canceling all the sounds and then a mechanism for  introducing {\it back} the desired sounds into the earphones. The latter, which is the focus of our work,  requires    programming  the output acoustic scene in real-time by semantically associating the individual incoming sounds with user input to determine which sounds to allow in the hearable device and which sounds to block.

\begin{figure}
    \centering
    \includegraphics[width=.4\textwidth]{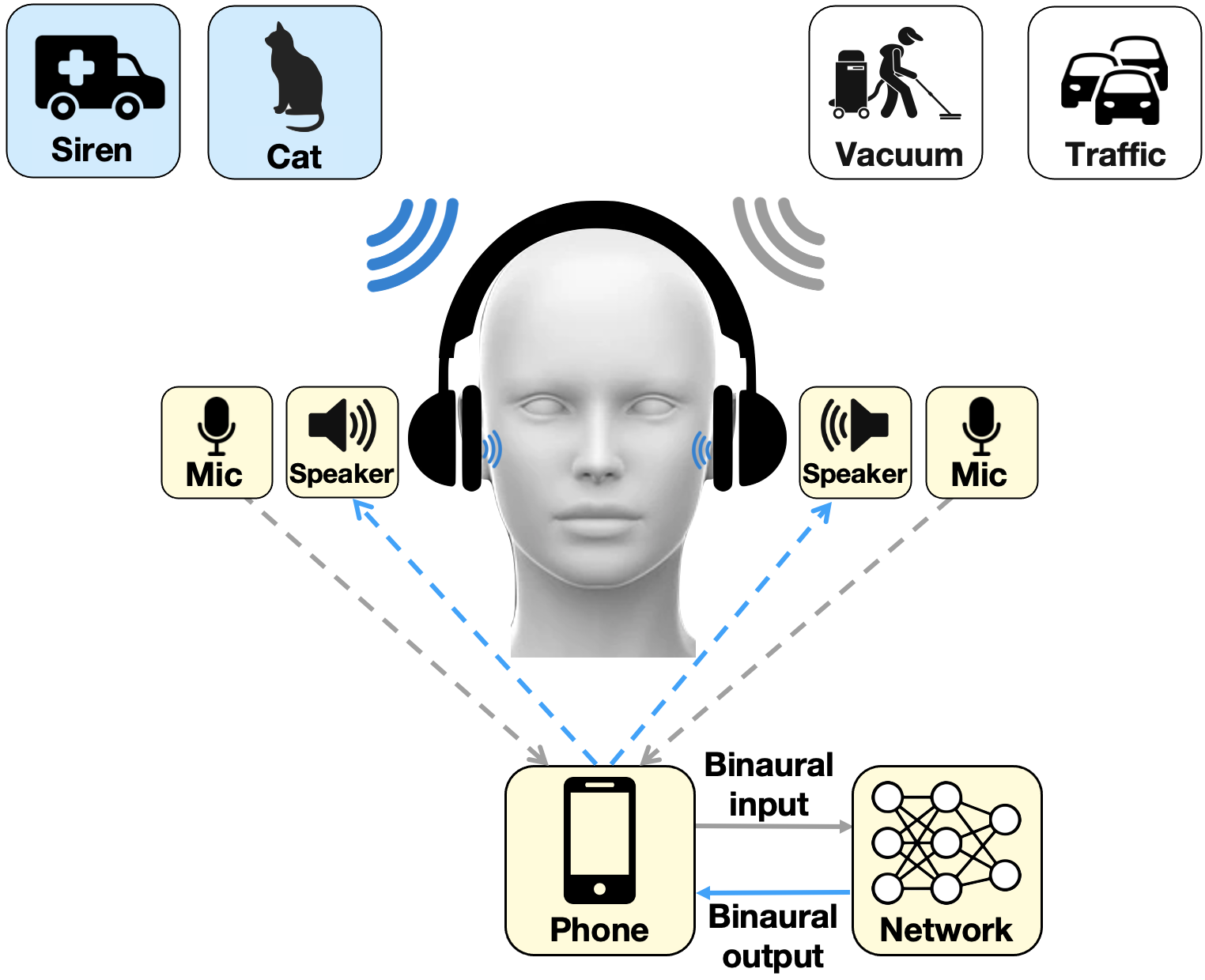}
    \caption{Semantic hearing architecture. The binaural input sounds are captured at a wired noise-canceling headset and sent to a phone, where we run on our sound extraction network. This  extracts the binaural output that captures the target sounds (e.g., sirens and cat sounds) and suppresses noise and interfering sounds (e.g., vacuum and traffic noise). This binaural output is played back  in real-time.}
  \vskip -0.15in
    \label{fig:arch}
\end{figure}

Animals have evolved over millions of years to  focus  on target   sounds and the associated directions~\cite{hearingnature}. However,   achieving this capability with in-ear devices like earphones and headsets is challenging for three key reasons.
\squishlist
    \item {\it Real-time requirements.} The sounds output by our design should be synced with the user's visual senses. This requires real-time  processing that satisfies stringent latency requirements. Research on medical hearing aids and augmented audio shows that we need a latency of less than 20-50~ms~\cite{stone1999tolerable,gupta2020acoustic}. This requires identifying the target sounds using 10 ms or less of audio blocks, separating them from interfering sounds, and then playing them back, all on a computationally-constrained device like a smartphone. 
    \item {\it Binaural processing.}  Sounds arrive at the two ears with different delays and attenuations~\cite{ieee}. The  physical separation between the two ears and the reflections/diffraction from the wearer’s head, i.e., the head-related transfer function,  provide
 cues for spatial perception. To preserve these cues, we need a binaural output to preserve or recover this spatial information for the target sounds across the two ears.  
 \item{\it Real-world generalization.}  While training and testing  a neural network on synthetic data is common in  audio machine learning research, designing a binaural target sound extraction  network  that generalizes to real-world hearable applications is challenging. This is because it is difficult to fully capture the complexity of real-world reverberations  and head-related transfer functions (HRTFs) in simulations.  We however require generalization to in-the-wild use in unseen acoustic environments across different users.


\squishend

In this paper, we address the above challenges and  demonstrate {\it semantic hearing}\footnote{Our inspiration for the name `semantic hearing' is directional hearing which is the ability to hear sounds from a specific direction~\cite{doclo2010acoustic,brayda2015spatially,directional}. Similarly,  semantic hearing is the ability to hear the sounds that are specified by some semantic descriptions, such as sound classes.} with  hearable devices. To achieve our goal, we make two key technical contributions.  We  design the first neural network capable of achieving binaural target sound extraction. Our network takes the two audio signals from the microphones at the two ears as binaural input and outputs two audio signals as binaural output, while preserving the directionality of the target sounds in the acoustic scene. To do  this, we start with our recent  single-channel (not binaural) transformer model for target sound extraction~\cite{waveformer}, which had neither real-world evaluation nor real-time smartphone operation. First, we optimize the network for real-time operations on smartphones. Then, we design a network that jointly processes the binaural input signals, allowing it to preserve the spatial information about the target sounds and output binaural audio (see~\xref{sec:network}). This joint processing is more effective at binaural target sound extraction and has half the computational cost of processing the binaural input signals separately.


\begin{figure}[t!]
\centering
\begin{subfigure}[t]{\linewidth}
    \includegraphics[width=\linewidth]{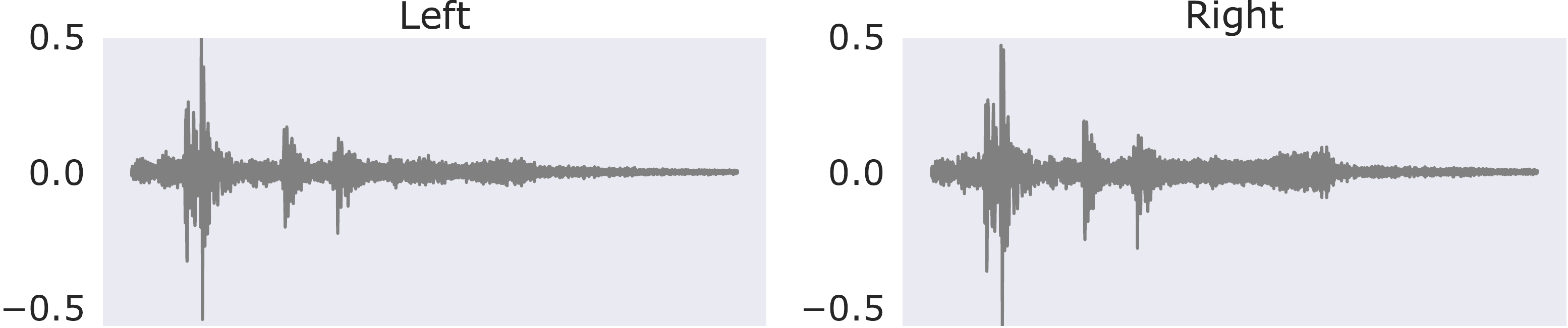}
    \label{fig:top_level}
    \vskip -0.15in
    \caption{Binaural input  of emergency  siren, chatter, and traffic noise.}
\end{subfigure}
\hfill
\centering
\begin{subfigure}[t]{\linewidth}
    \includegraphics[width=\linewidth]{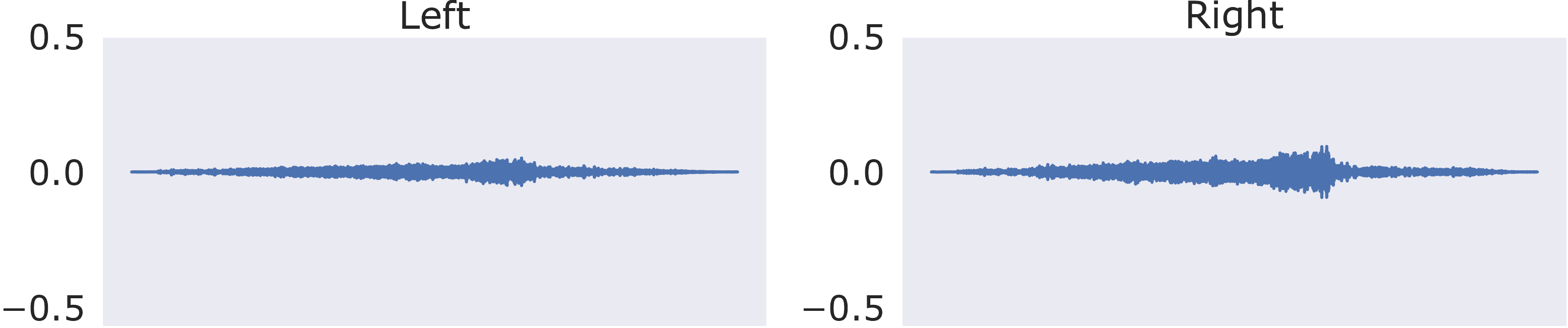}
    \label{fig:enc}
    \vskip -0.15in
    \caption{Binaural output with siren extracted.}
\end{subfigure}
\begin{subfigure}[t]{\linewidth}
    \includegraphics[width=\linewidth]{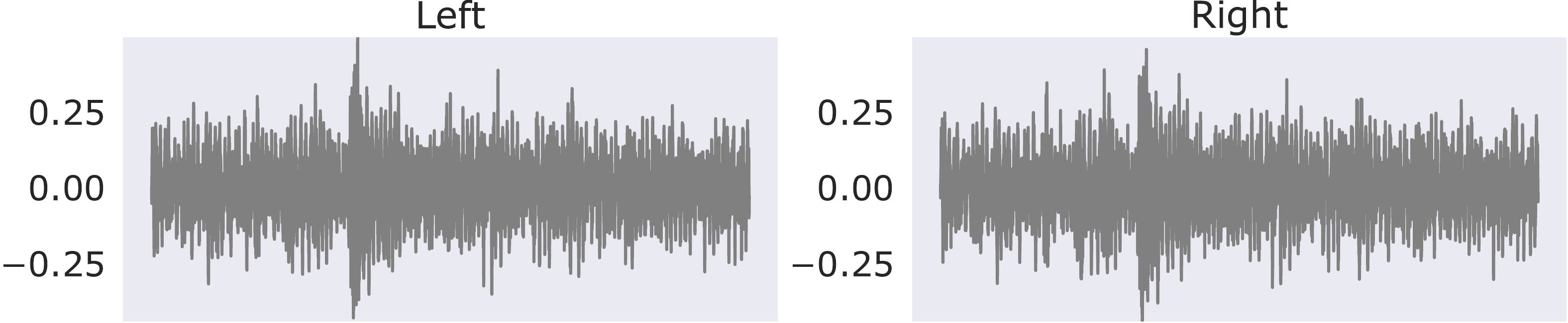}
    \label{fig:top_level}
    \vskip -0.15in
    \caption{Binaural input of birds chirping, chatter, and street noise.}
\end{subfigure}
\hfill
\centering
\begin{subfigure}[t]{\linewidth}
    \includegraphics[width=\linewidth]{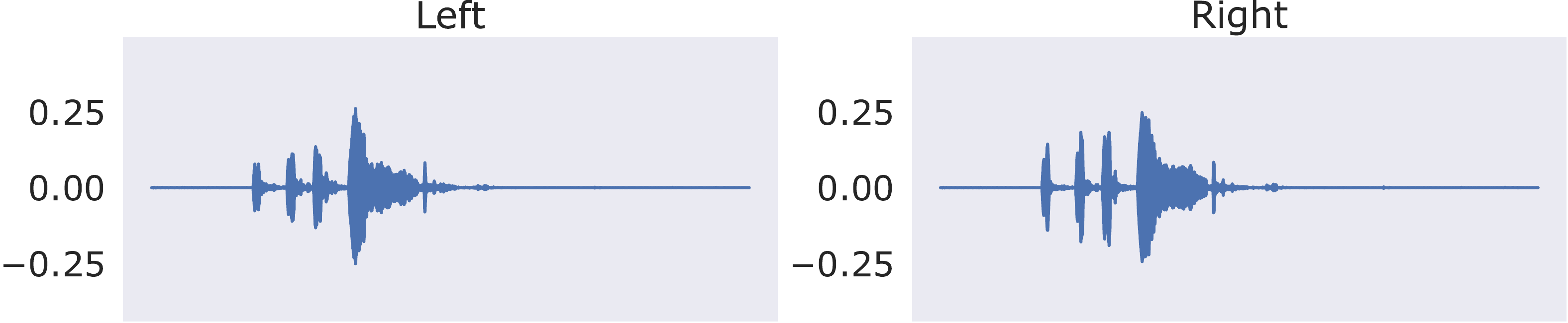}
    \label{fig:enc}
    \vskip -0.15in
    \caption{Binaural output with birds chirping extracted.}
\end{subfigure}
    \vskip -0.1in
    \caption{Real-world  binaural input and output recordings obtained with our semantic hearing system.}
    \vskip -0.15in
    \label{fig:qual_demo}
\end{figure}

We also design a training methodology that ensures our binaural network can generalize to real-world situations, such as reverberations, multipath, and HRTFs. Obtaining training data in fully natural environments can be difficult because we may capture mixtures but lack access to the ground truth sounds needed for supervised learning. Moreover, training a network that can generalize to in-the-wild use with hearables requires that the training data capture reverberations, multipath, and head-related transfer functions across a large number of users. To achieve this, we synthesize our training data using multiple datasets. First, we use an HRTF dataset, which includes measurements from 41 users in  non-reverberant environments. We convolve the room impulse responses with thousands of examples from 20 different audio classes to generate both our mixtures and the  ground truth binaural audio. However, this does not capture the reverb and multipath in realistic environments. Therefore, we augment these synthesized mixtures with training data synthesized from three different datasets that provide binaural room impulse responses captured in real rooms.  This facilitates our network to generalize to users and real-world environments that are not in the training dataset.

To demonstrate proof-of-concept, we  augmented an off-the-shelf noise-canceling headset with commercial wired binaural earphones that provide access to data from both microphones. We  implement our neural network on a connected smartphone and train it with 20 different sound classes, including sirens, baby cries, speech, vacuum cleaners, alarm clocks, and bird chirps. Our results are as follows.

\aptLtoX[graphic=no,type=html]{\begin{itemize}}{\squishlist}
\item We achieve an average signal improvement of  7.17~dB  across the 20 target sounds, in the presence of interfering sounds and urban background noise. Our real-time network has a 6.56~ms runtime on  iPhone 11 for processing  a 10~ms chunk of binaural audio. 
\item In-the-wild evaluation with  participants in various indoor and
outdoor scenarios with our hardware shows that our system can extract the target sounds (Fig.~\ref{fig:qual_demo}) and generalize to  previously unseen participants,  and environments, without requiring any training data collection with our hearable hardware.
\item In a spatial hearing study where we played sounds from different directions  in  five previously unseen rooms, participants were able to predict the direction of the target sounds output by our system with 50th and 90th percentile errors of $22.5^\circ$ and
$45^\circ$ respectively. These errors were similar for noise-free clean sounds.


\item In a user study with 22 participants who spent over 330 minutes rating binaural data from real-world indoor and outdoor environments, our system achieved a higher mean opinion score and interference removal for the target sounds  than  the binaural input. 
\aptLtoX[graphic=no,type=html]{\end{itemize}}{\squishend}


\vskip 0.05in\noindent{\bf Contributions.} We introduce the concept of semantic hearing, where we can program the binaural acoustic scene based on semantic sound descriptions. Our work makes five  key contributions. 1) we present the first neural network to achieve binaural target sound separation and demonstrate that our network can run in real-time on smartphones, 2) we design a training methodology to generalize our system to unseen real-world environments, and users, 3) we implement a proof-of-concept with off-the-shelf hardware and show that our system achieves the above goals in real-world environments, 4) we highlight where our current system fails and opportunities for future research, and 5) by making our binaural models and datasets public, we hope to kickstart future research in the community towards further developing the concept of semantic hearing in practical hearable applications.


\section{Background and Related Work}

\begin{figure}
    \centering
    \includegraphics[width=.4\textwidth]{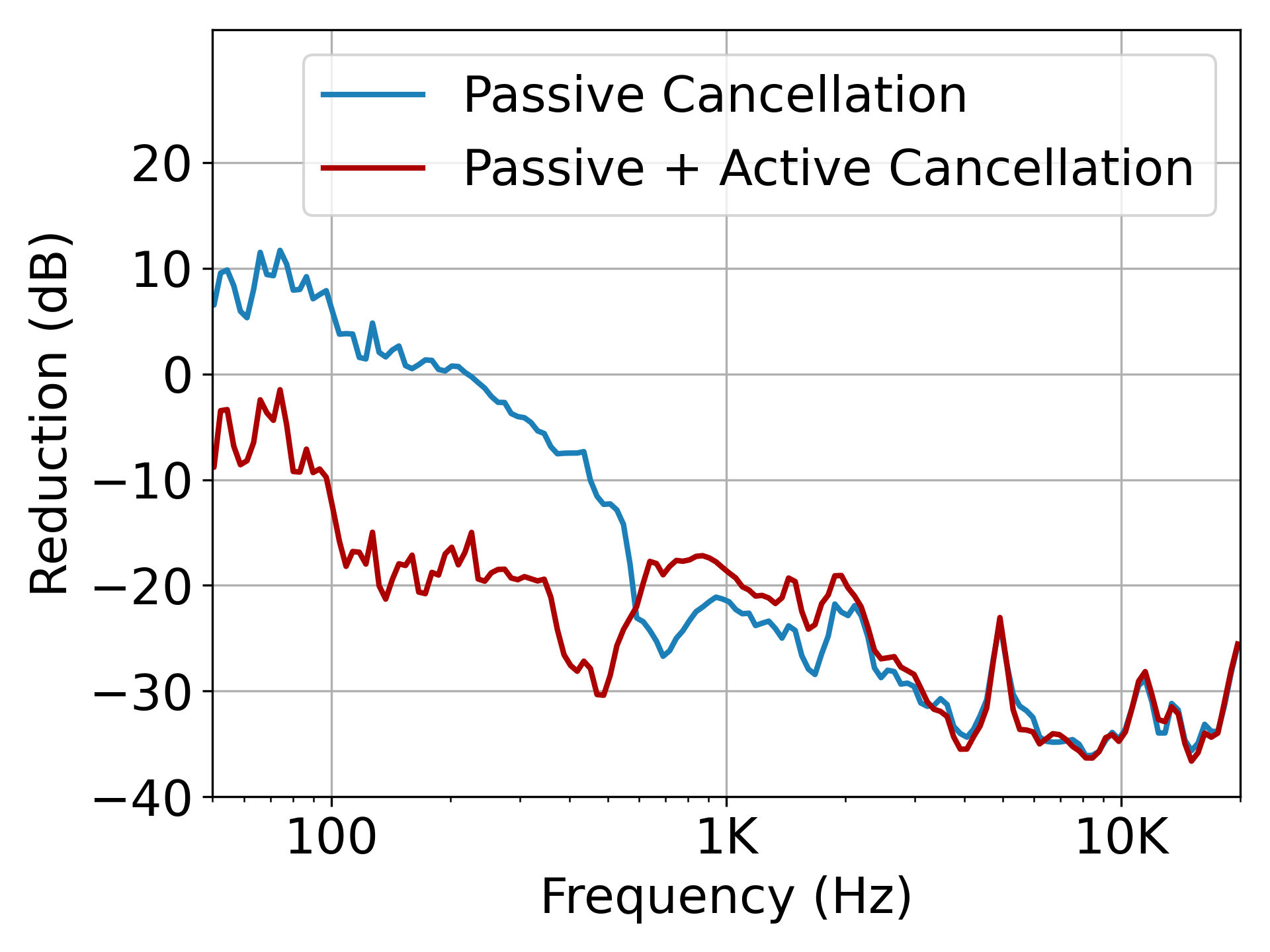}
    \vskip -0.15in
    \caption{{Noise reduction achieved with Sony WH-1000XM4 headphones -- with and without active noise cancellation turned on -- measured using an in-ear microphone inside the headphone cup. The reduction is measured relative to a microphone recording outside the ear cup. The spuriously large values at low frequencies (< 100~Hz) are due to the in-ear microphones picking up the wearer's blood pulse.}}
  \vskip -0.15in
    \label{fig:sonyxm4}
\end{figure}

Over the last decade, noise-canceling headsets and earbuds have undergone significant improvements, which now allow for more effective attenuation of {\it all} sounds in the environment. {In fact, our  experiments, where we play white noise to a human subject wearing a pair of Sony WH-1000XM4 headphones, show the impressive attenuation capabilities of these modern systems (Fig.~\ref{fig:sonyxm4}).}  We identify this as an opportunity that provides us with an acoustic clean slate to introduce back target binaural sounds of interest from the environment.  To the best of our knowledge, none of the prior work has  explored semantic hearing capabilities for hearables. In the rest of this section, we describe   related work in hearable systems,    signal processing and  machine learning for audio,   and interaction tools.

{\bf Active noise cancellation and acoustic transparency.}  Active noise cancellation is a well-studied problem where  outward-facing microphones are used to  capture sounds~\cite{cancel}. An anti-noise signal is then  transmitted  to cancel {\it all}  the external sounds and noise, which has more stringent delay requirements than semantic hearing. Traditional noise cancellation systems required bulky headsets. However in recent years lightweight {in-ear} earbud systems like the AirPods Pro  can achieve reasonable noise-cancellation in many practical scenarios~\cite{airpods}. Semantic hearing leverages noise-cancelling earphones to cancel {\it all} sounds and then uses the mechanisms in this paper to program  acoustic scenes in real-time.

The acoustic transparency mode for in-ear devices tries to imitate the sound response of an open-ear system by transmitting the appropriate signals into the ear canal~\cite{transparency2}. Like active noise cancellation, this is agnostic to the sound classes. Adaptive transparency on Apple airpods is designed to  automatically reduce  the amplitude of loud sounds~\cite{adaptivetransparency}. While related, this does not allow the user to pick and choose which sound classes to hear.

{\bf  Speech systems.}  Prior systems have predominantly focused on improving the performance of speech-related tasks for in-ear devices (e.g., Airpods), telephony (e.g., Microsoft Teams), and voice assistants (e.g., Google Home). This includes speech enhancement~\cite{chatterjee2022clearbuds, jon}, target speech extraction~\cite{eskimez2022personalized, giri2021personalized}, and speech separation~\cite{subakan2021attention, luo2022tiny}. Oftentimes, these systems collectively regard all non-speech sounds just as noise. In contrast, semantic hearing requires understanding the semantics of various natural and artificial sounds in real-time, in the presence of interfering sounds, and determining which sounds to allow and which to block, based on user input. Speech is one amongst many other sound classes  in our system. 

\begin{figure}[t]
    \centering
    \includegraphics[width=.49\textwidth]{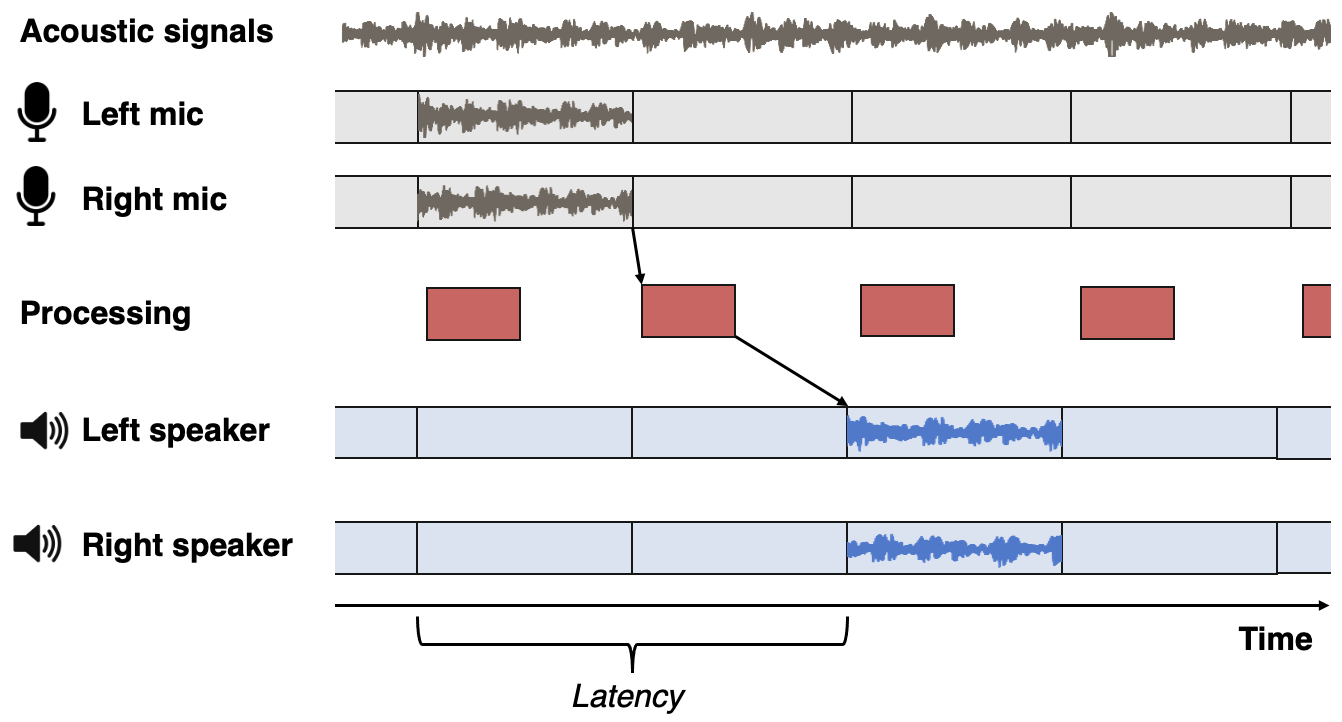}
    \vskip -0.15in
    \caption{System requirements.  Different components that contribute to latency in binaural target sound extraction. }
   \vskip -0.15in
    \label{fig:latency}
\end{figure}

{\bf Neural networks for target sound extraction.}  Target sound extraction is the task of  separating one or a limited number of target sounds  from a mixture of sounds.  Compared with speech systems, this is an underexplored problem in the  audio machine learning community. However recent works have proposed neural networks that can achieve target sound extraction where clues about the target sound are provided either via audio~\cite{delcroix2022soundbeam,gfeller2021one},  images~\cite{gao2019co,xu2019recursive},  text~\cite{kilgour2022text,liu2022separate}, onomatopoeic words~\cite{okamoto2022environmental}, or a one-hot  vectors~\cite{2020arXiv200605712O}. All these models are designed for  offline processing of audio clips, where the neural network has access to the entire audio file  ($\ge$ 1 s) and hence cannot support our real-time hearable use-case. 

The closest related work is our recent research on Waveformer \cite{waveformer}, which introduces a neural network architecture for target sound extraction. Waveformer was shown to run in real-time on a desktop computer. Our work differs from \cite{waveformer} in two important dimensions. First, Waveformer is a single-channel model that operates on a single microphone. In contrast, our target use-case requires binaural processing across the two ears. As we show in~\xref{sec:benchmark}, running the prior model independently on the two microphones is computationally expensive, failing to meet the real-time requirements on a smartphone. Second, all prior work in this domain was evaluated on synthetic datasets and has not been demonstrated on hardware in real-world scenarios. In contrast, we present the first binaural target sound extraction system that can run in real-time on smartphones. We designed a training methodology that allows our system to generalize to unseen indoor and outdoor real-world environments.


{\bf  Hearable applications.}  Recent work has used in-ear sensors for health applications~\cite{tam1, infection, oae} and activity tracking~\cite{mobisys21,teeth}. Prior work has also explored various interaction modalities like ultrasound sensing~\cite{chi22-ultrasonic} and on-face interaction~\cite{earbuddy} for in-ear devices. The closest to our work is Clearbuds~\cite{chatterjee2022clearbuds}, which focuses on the task of enhancing the speech of the wearer using synchronized audio signals from two wireless earbuds. This prior work is focused on speech enhancement and is complementary to our system. Further, since the target application for \cite{chatterjee2022clearbuds} is telephony, it uses a 44.8~ms lookahead and has a latency of 109~ms. 

{\bf Audio-based tools.}  Prior work has explored the use of  sounds to perform activity  recognition for wearables and smart home applications~\cite{bodyscope,soundsense,samosa, ubicoustics,tonami2022sound,dhruv,dhruv2}.  These systems operate on around 1s audio chunks as the target use cases do not have the O(10~ms) latency requirements of in-ear audio applications.  Prior work has also designed interaction tools for audio editing~\cite{editing1,editing2}. Our work is complementary in that it is focused on in-ear audio applications and semantic hearing that has  more stringent latency requirements.

\section{Semantic hearing}
We first describe our system requirements and  then present the network architecture we use for real-time binaural target sound extraction  on smartphones. Next, we present our training methodology that generalizes our design  to real-world  use. 

\subsection{System Requirements} \label{requirements}



The goal of our design is to program the acoustic environment with imperceptible latency such that the target sound of interest is present but all other interfering  sounds are suppressed.  Given the stringent latency constraints, we cannot perform the necessary computation in the cloud but have to operate in real-time using computationally constrained devices like  smartphones.  Further, the target sounds generated by the model must originate from the same spatial directions as the real-world target sounds. Thus, our design must meet two key requirements: 1) real-time low-latency operation, and  2) binaural real-world generalization.

\vskip 0.05in\noindent{\it Real-time low-latency operation.} Fig.~\ref{fig:latency} shows the different components that contribute to end-to-end latency in binaural acoustic processing systems.  The first step is to feed  the sound signals into  two memory buffers of the binaural microphones. The acoustic data  from the two microphones in each block is then fed into our neural network that outputs a block-length worth of binaural target sound data. This binaural output is then played back through the two speakers on the headset.  

To ensure that the audio played through the headset is synced with the user's visual senses, we need this end-to-end latency to be less than 20-50~ms~\cite{stone1999tolerable,gupta2020acoustic,directional}. To achieve this, we need to reduce the buffer duration, the look-ahead duration and the processing time. This is challenging for multiple reasons. 1) A small buffer duration  of say 10~ms means that the algorithm has only an 10~ms block of current data to not only understand the semantics of the acoustic scene but also separate the target sound  from other interfering sounds. While we can use the acoustic signals that arrived prior to the current block,  many of our target sounds (e.g., door knocks) are not continuous. Reducing the buffer size even further to say 2~ms can be challenging from an operating system perspective since it can increase the number of system calls. 2) While a large lookahead can provide more context for the neural network to extract the target sounds, meeting our end-to-end latency requirement reduces the leeway we have in terms of the available lookahead to a few milliseconds. 3) Real-time operation requires processing each acoustic block within the duration of the block itself. This means that it should take less than 10~ms to process a 10~ms buffer~\cite{directional}. This can be  challenging since neural networks are not known for their lightweight computation. Further, since we cannot send the data to the cloud, the processing must be performed on-device on computationally-constrained devices like smartphones. In addition to  all the above constraints, the operating systems also has I/O delays which for audio  on iOS is on the order of 4~ms, depending on the buffer size~\cite{ioslatency}. 

\vskip 0.05in\noindent{\it Binaural real-world generalization.} In real life,  the target sounds experience reverberations and multipath propagation due to reflections from walls and other objects in the environment. Further, the human head and torso reflect and obstruct sounds. As a result the target sound arrives with different amplitudes and delays at the two ears. The differences in the received sounds across the two ears provide spatial awareness to humans. Thus, it is critical in our design to preserve these differences and play the target sounds with different amplitudes and delays through the two speakers of the headset. This is challenging since  the target and interfering sounds can be  at different positions and experience different reverberations and reflections from the  head-related transfer function. Further, the multipath effects and reverberations are difficult to predict in real-world environments, let alone the fact that the head-related transfer functions can change across wearers.

\begin{figure*}[t!]
\centering
\begin{subfigure}[t]{0.26\textwidth}
    \includegraphics[width=\linewidth]{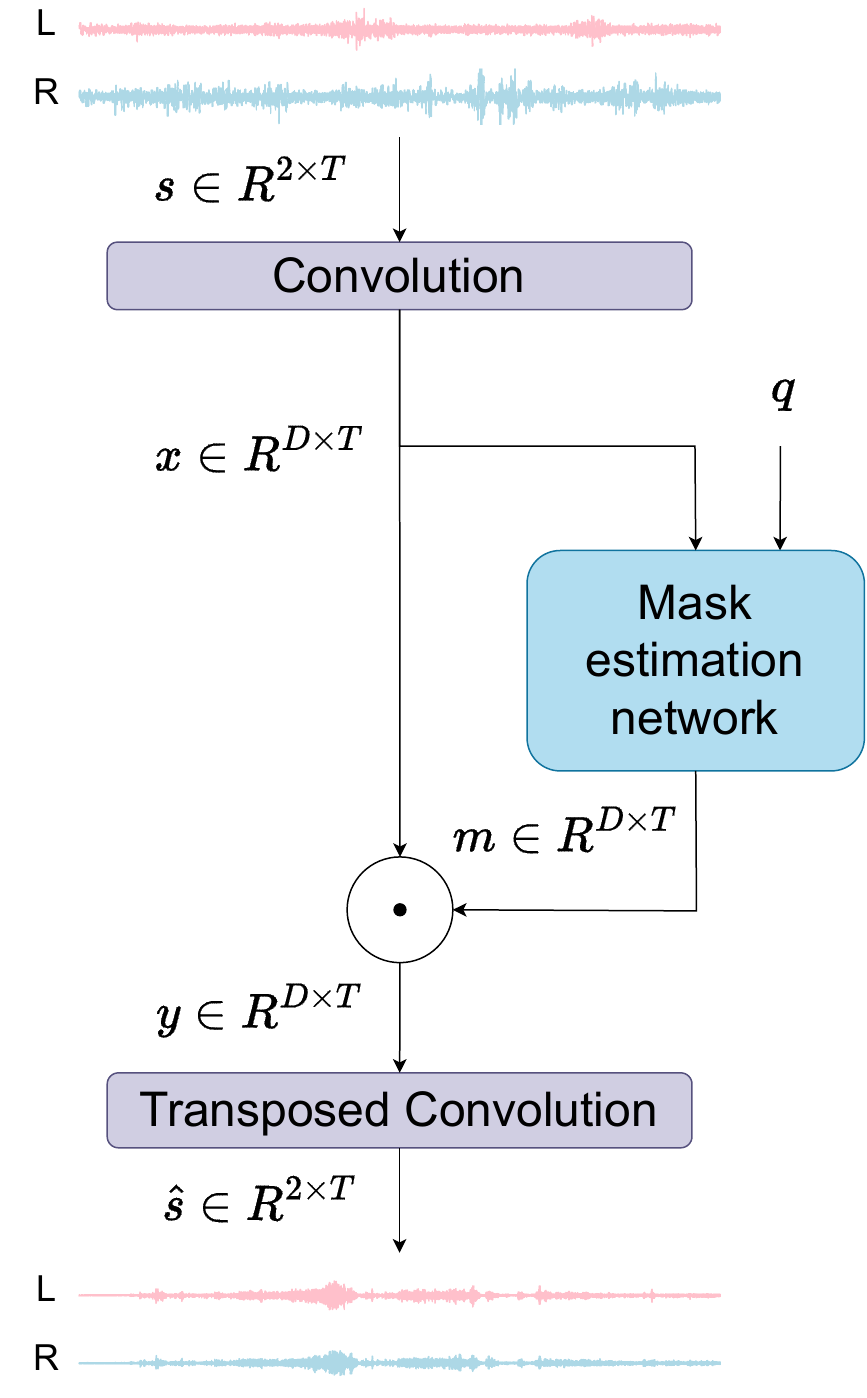}
    \label{fig:top_level}
    \vskip -0.15in
    \caption{Binaural extraction framework}
\end{subfigure}
\hfill
\centering
\begin{subfigure}[t]{.44\textwidth}
    \includegraphics[width=\linewidth]{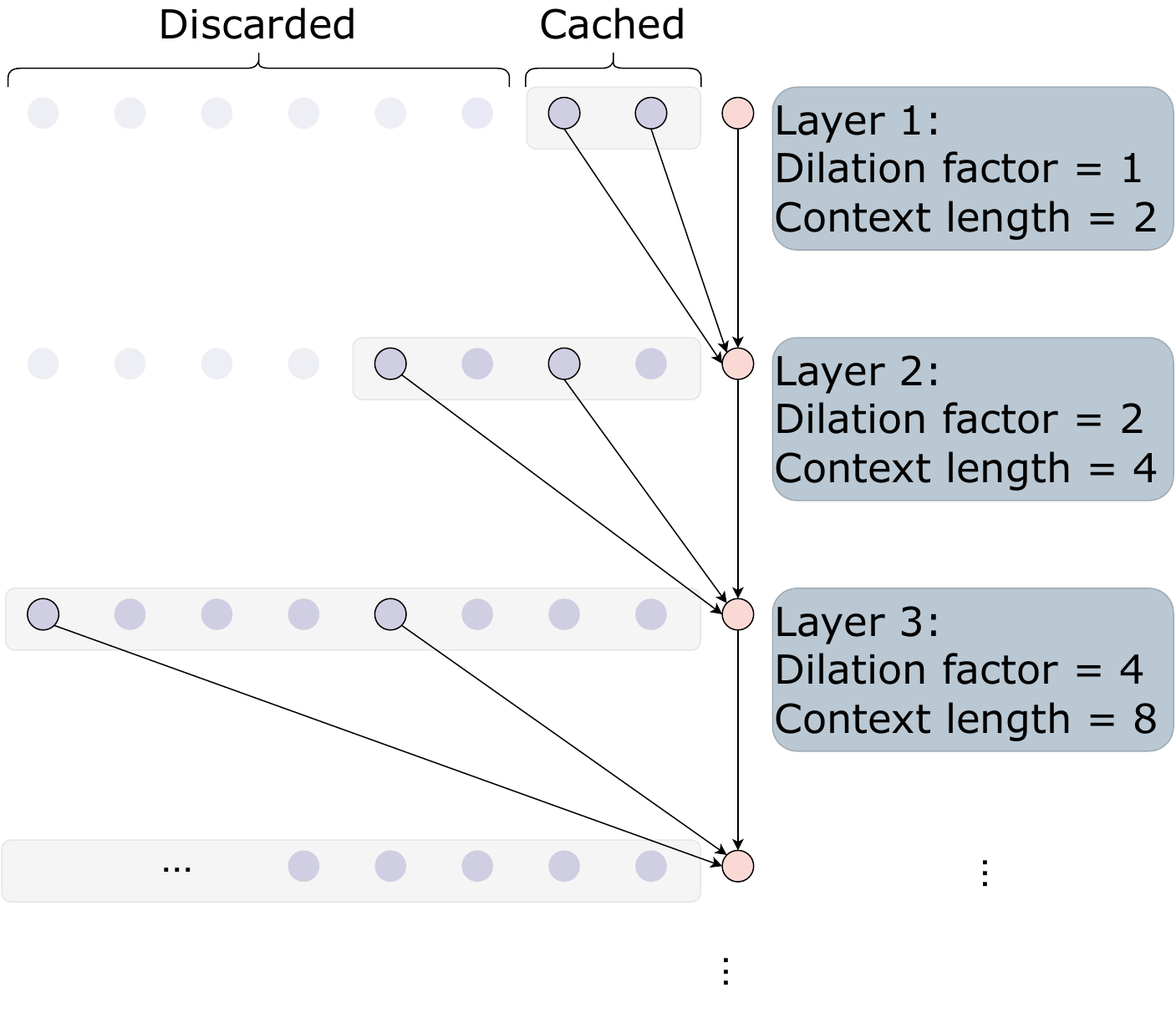}
    \label{fig:enc}
        \vskip -0.15in
    \caption{Encoder}
\end{subfigure}
\hfill
\centering
\begin{subfigure}[t]{.15\textwidth}
    \includegraphics[width=\linewidth]{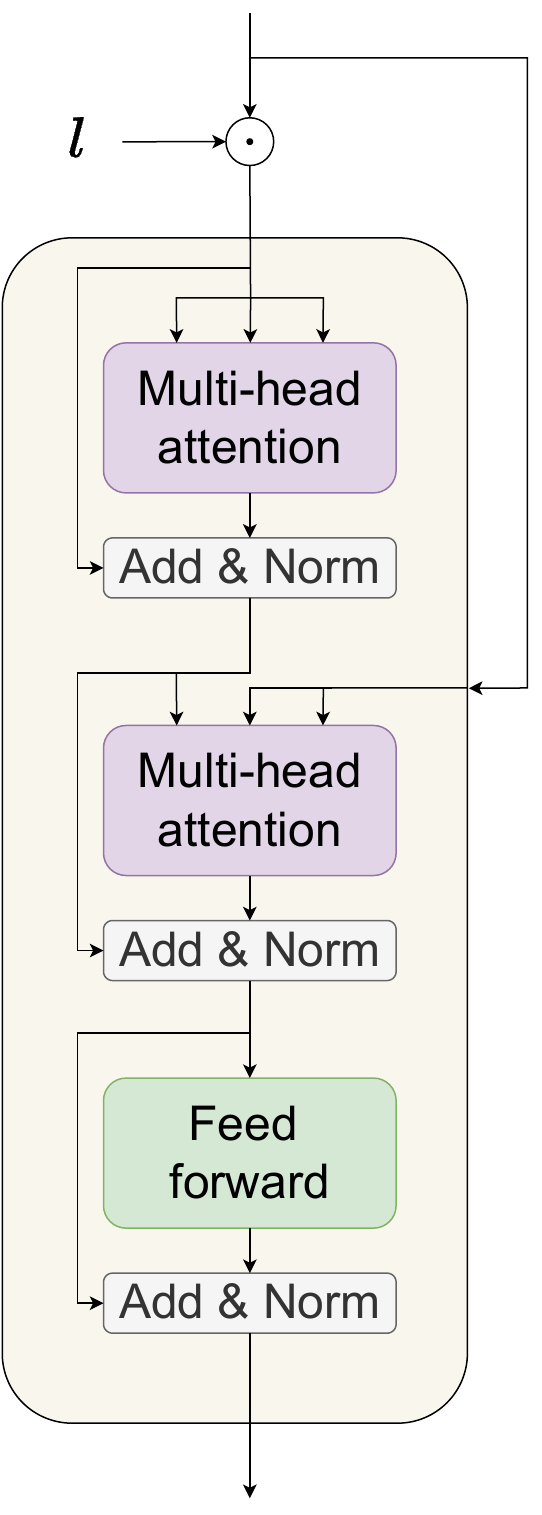}
    \label{fig:dec}
        \vskip -0.15in
    \caption{Decoder}
\end{subfigure}
\vskip -0.1in
    \caption{Binaural target sound extraction network architecture. a) Our high-level binaural extraction framework. Mask estimation network is an encoder-decoder architecture operating on latent space representation of binaural signals to extract mask for target sound based on the query vector $q$. b) and c) show the encoder and decoder architectures used in the mask estimation network. The encoder processes the previous input context and does not consider the label embedding. Decoder first conditions the encoded representation with the label embedding, $l$, and then generates the mask corresponding to the target sound using the conditioned representation.}
    \label{fig:network}
\end{figure*}


\subsection{Binaural target sound extraction network}\label{sec:network}

We first explain the high-level framework for our binaural target sound extraction neural network. Then we explain the causal and streaming adaptation of this network. Finally, we provide a detailed description of the network architecture.

\subsubsection{High-level framework} \label{framework} Consider $s \in R^{2 \times T}$ to be the input binaural signal provided to the target sound extraction network. {Since time-domain models also have been shown to be able to learn representations analogous to STFT features~\cite{luo2019conv}, our network operates on  time-domain binaural signals.} As shown in  Fig.~\ref{fig:network}a, the signal is first mapped to a representation in a latent space, $x \in R^{D \times (T / L)}$, by using a 1D convolution layer with a kernel size $\geq$ $L$ and a stride equal to $L$. $D$ and $L$ are tuneable hyperparameters of the model. $D$ is the dimensionality of the model, having a significant effect on the parameter count, and consequently the computational and memory complexities. $L$ determines the duration of the smallest audio chunk that can be processed with the model. The latent space representation $x$, is then passed to a mask generator, $\mathcal{M}$, which estimates an element-wise mask $m$ as:
\begin{equation}
    m = \mathcal{M}(x, q) \ | \  m \in R ^{D \times (T / L)}; \ q \in \{0,1\}^{N_c},
\end{equation}
where $N_c$ is the total number of sound classes the model is trained for. The representation corresponding to the target sound is obtained by element-wise multiplication of the input representation, $x$, and the mask, $m$, as follows:
\begin{equation}
    y = x \odot m \ | \ y \in R^{D \times (T / L)}.
\end{equation}
The output audio signal $\hat{s} \in R^{2 \times T}$ is then obtained by applying a 1D transposed convolution on $y$, with a stride of $L$.

In contrast to more complex binaural extraction frameworks proposed specifically for {\it speech}, where each channel is separately  and parallely processed \cite{han2020realtime, gu2019endtoend}, our design jointly processes the two channels for computational efficiency.  In our experiments, we show that our simpler framework performs competitively with the prior parallel processing frameworks  in terms of target sound extraction accuracy, even with a 50\% lower runtime cost.


\subsubsection{Streaming inference and causality} \label{streaming}
For real-time on-device operation, the model must output the audio corresponding to the target sound as soon as the input audio is received, i.e., within the latency requirements detailed in  \xref{requirements}. Since the audio is fed to the model from  the device buffers, the buffer size determines the duration of the audio chunk the model receives at each time step. Assuming the buffer size to be divisible by the stride size $L$, the audio chunk size can be represented as the number of strides, $K$. That is, the buffer size of an audio chunk of size $K$ is equal to $KL$ samples. Such a real-time setup means that the model only has access to the current and previous chunks, but not future chunks. \emph{This requires the model to be causal with the time resolution of the buffer size, i.e., $KL$ audio samples}. As a result, in the high-level framework described above, the input convolution, the mask estimation block, the element-wise multiplication, and the output transposed convolution must operate on one audio chunk at each time step.

The binaural target sound extraction framework described in  \xref{framework} can be adapted to chunk-wise streaming inference as follows. Consider the input audio signal corresponding to the $k$th chunk to be $s_k \in R^{2 \times KL}$. The input 1D convolution maps this audio chunk to its latent space representation, $x_k \in R^{D \times K}$. The mask estimation block is then used to estimate the mask corresponding to the target sound, based on the current chunk, as well as a finite number of the previous chunks:
\begin{equation}
    m_k = \mathcal{M}(x_k, q, x_{k-1}, x_{k-2}, ...) \  | \  m_k \in R^{D \times K}. 
        \label{eq:mask_est}
\end{equation}
The previous chunks act as the audio context for the neural network, referred to as the receptive field of the model. A receptive field of 1-1.5s is shown to result in good performance \cite{luo2019conv}. The output representation of the current chunk corresponding to the target sound, $y_k \in R^{D \times K}$ can then be obtained as:
\begin{equation}
    y_k = x_k \odot m_k. 
\end{equation}
The resulting output representation is then converted to the output signal $\hat{s}_k \in R^{2 x KL}$ by applying the 1D transposed convolution. 

\subsubsection{Mask estimation network} \label{mask_est} Several architectures have been proposed in the literature for mask estimation such as Conv-TasNet \cite{luo2019conv}, U-Net \cite{Jansson2017SingingVS}, SepFormer \cite{subakan2021attention}, ReSepFormer \cite{subakan2022resource}, and Waveformer \cite{waveformer}. 
Waveformer is an recently proposed efficient streaming architecture implementing chunk-based processing, which makes it suitable for our task. In this work, we propose a modified version of Waveformer to further increase efficiency without any loss in performance. The mask estimation network is an encoder-decoder neural network architecture, where the encoder is purely convolution-based and the decoder is a transformer decoder.

Different from Waveformer, in this work, we use the same dimensionality for both the encoder and decoder. This allowed us to use the standard transformer decoder \cite{https://doi.org/10.48550/arxiv.1706.03762}, instead of a modified one used in the Waveformer. Waveformer proposes a smaller dimensionality for the decoder block, compared to the rest of the model. The transition between different dimensionality is achieved using projection layers (1D convolution layers with kernel size equal to 1). This however breaks the residual paths and the result might affect the gradient flowing back, which is mitigated in the Waveformer using a long residual connection bypassing the decoder. For our binaural  application, however, we found that different dimensionality is not necessarily providing gains warranting the complexity of the projection layers and the long residual connection. 

\vskip 0.05in\noindent\textbf{Encoder.} Mask estimation in Eq.~\ref{eq:mask_est}, involves processing many previous chunks in addition to  the current chunk to obtain the mask corresponding to the current chunk. Repeated processing of the entire receptive field for each iteration could become intractable for a real-time on-device application. To mitigate this inefficiency, while achieving a large receptive field, our mask estimation network implements a Wavenet \cite{https://doi.org/10.48550/arxiv.1609.03499} style dilated causal convolutions for processing the input and previous chunks. In this work, for efficient on-device inference, we implemented the dynamic programming algorithm proposed in Fast Wavenet \cite{paine2016fast}. As shown in  Fig.~\ref{fig:network}b, higher  efficiency is achieved by reusing the intermediate results computed in the previous iterations. The encoder function $\mathcal{E}$ processes the input chunk $x_k$ and an encoder context $\xi_k$ to generate the encoded representation of the input chunk:
\begin{equation}
    e_k, \xi_{k + 1} = \mathcal{E} (x_k, \xi_{k}) \ | \  e_k \in R^{D \times K}
\end{equation}
The size of the context $\xi_k$ depends on the hyperparameters of the encoder. In our implementation, the encoder is comprised of a stack of 10 dilated causal convolution layers. The kernel size of all layers is equal to 3, and the dilation factor is progressively doubled after each layer starting with 1, resulting in dilation factors $\{2^0, 2^1, ..., 2^9\}$. Since the kernel size is equal to 3, the context needed for each dilated convolution layer is twice the layer's dilation factor. As long as this context is saved after each iteration, and padded with the input chunk in the next iteration, the intermediate results corresponding to the previous chunks do not have to be recomputed. Thus the size of the context $\xi_k$  is equal to $2 \times \sum_{i=0}^{9} 2^i = 2046$.

\vskip 0.05in\noindent\textbf{Decoder.} The query vector $q$ is first embedded into the embedding space using a linear layer to generate a label embedding $l \in R^D$. The mask corresponding to the target sound $m_k$ is estimated using a transformer decoder layer \cite{https://doi.org/10.48550/arxiv.1706.03762}, represented here as $\mathcal{D}$. The encoded representation is first conditioned with the label embedding $l$ by an element-wise multiplication. The encoded representation and the conditioned encoded representation are first concatenated in the time dimension, with those from the previous time step, before processing with the transformer decoder layer $\mathcal{D}$. The encoded representation from the previous time step, $e_{k-1}$, acts as the decoder context. The mask estimation can be written as:
\begin{equation}
    m_k = \mathcal{D} (\{l\cdot e_{k-1}, l\cdot e_k\}, \{e_{k-1}, e_k\}, )
\end{equation}
where $\{\}$ represents concatenation in the time dimension. As shown in Fig.~\ref{fig:network}c, the transformer decoder $\mathcal{D}$ first computes the self-attention result of the conditioned encoded representation $\{l\cdot e_{k-1}, l\cdot e_k\}$ using the first multi-head attention block, followed by cross-attention between the self-attention result and the unconditioned encoded representation $\{e_{k-1}, e_k\}$ using the second multi-head attention block. A feed-forward block along with residual connection generates the final mask corresponding to the target sound.



\begin{table*}[t!]
\centering
\caption{Number of raw audio files collected for training/testing/validation from each dataset for  our potential target classes. { The total number of mixtures we generate using these for training, testing and validation are 100k, 10k, and 1k, respectively.}}\label{tab:table1}
\vskip -0.1in
{\footnotesize\tabcolsep5.3pt
\begin{tabular}{lcccccccccc}
\toprule
Dataset & Alarm clock &  Baby cry & Birds chirping & Car horn &      Cat & Rooster crow & Typing &  Cricket &       Dog & Door knock \\
\midrule
 FSD50K &      34/4/4 &    30/8/4 &        65/52/8 &  36/21/4 &  73/39/9 &            23/9/3 &         78/68/9 &  61/14/7 & 109/33/13 &    65/33/8 \\
 ESC-50 &      24/8/8 &    24/8/8 &         24/8/8 &   24/8/8 &   24/8/8 &            24/8/8 &          24/8/8 &   24/8/8 &    24/8/8 &     24/8/8 \\
MUSDB18 &       0/0/0 &     0/0/0 &          0/0/0 &    0/0/0 &    0/0/0 &             0/0/0 &           0/0/0 &    0/0/0 &     0/0/0 &      0/0/0 \\
  DISCO &       0/0/0 &  95/53/11 &          0/0/0 &    0/0/0 &    0/0/0 &             0/0/0 &           0/0/0 &    0/0/0 &     0/0/0 &      0/0/0 \\
  Total &    58/12/12 & 149/69/23 &       89/60/16 & 60/29/12 & 97/47/17 &          47/17/11 &       102/76/17 & 85/22/15 & 133/41/21 &   89/41/16 \\
\bottomrule
\end{tabular}
\begin{tabular}{lcccccccccc}
\toprule
Dataset & Glass breaking &   Gunshot &   Hammer &      Music &     Ocean &    Singing &    Siren &     Speech & Thunderstorm & Toilet flush \\
\midrule
 FSD50k &      212/31/24 & 169/67/20 & 88/39/10 &      0/0/0 &  86/19/10 &  134/76/19 &   16/5/3 & 494/109/56 &     122/9/14 &    112/21/13 \\
 ESC-50 &         24/8/8 &     0/0/0 &    0/0/0 &      0/0/0 &    24/8/8 &      0/0/0 &   24/8/8 &      0/0/0 &       24/8/8 &       24/8/8 \\
MUSDB18 &          0/0/0 &     0/0/0 &    0/0/0 & 581/358/62 &     0/0/0 & 480/291/54 &    0/0/0 &      0/0/0 &        0/0/0 &        0/0/0 \\
  DISCO &          0/0/0 &     0/0/0 &    0/0/0 &      0/0/0 &     0/0/0 &      0/0/0 &    0/0/0 &      0/0/0 &        0/0/0 &        0/0/0 \\
  Total &      236/39/32 & 169/67/20 & 88/39/10 & 581/358/62 & 110/27/18 & 614/367/73 & 40/13/11 & 494/109/56 &    146/17/22 &    136/29/21 \\
\bottomrule
\end{tabular}
}
\end{table*}\setcounter{table}{1}

\subsection{Training for real-world generalization} \label{training}

We first describe our audio class dataset curation and then present our training methodology to generalize to real-world  scenarios.

\subsubsection{Picking audio classes.} Our main goal is to create a system that efficiently handles target sounds encountered in real-world situations. By focusing on practical applications, we  identify a manageable set of target sound classes to extract. However, in reality, we come across a wide range of background sounds, many of which are not part of our target sound classes.  To curate our dataset of sound classes, we follow the AudioSet ontology~\cite{audioset}, which provides a comprehensive and structured representation of the relationships between various sound classes. The ontology arranges the sound classes as nodes in a graph and groups them into seven main sound categories. Each sound class node has a unique AudioSet ID and may contain one or more child nodes that represent more specific sound classes. For example, the ``Hands'' sound class  has two children, namely ``Finger Snapping''  and ``Clapping.'' In the rest of this section, we describe how we pick our target sound classes as well as the interfering classes.

\aptLtoX[graphic=no,type=html]{\begin{itemize}}{\squishlist}
\item {\bf Target sound classes.} We first consider various indoor and outdoor scenarios where the system is likely to operate, such as beaches, parks, streets, living rooms, offices, and cafes. Based on these scenarios, we identify potential sound sources that are prevalent in such locations, such as human speech, dogs, cats, birds, sea waves, and music. We then compile a list of sound classes associated with these selected target sounds and map each of these classes to a label in the AudioSet ontology. We eventually selected 20 sound classes that we felt human listeners could distinguish with reasonably high accuracy.  


\item {\bf Other sound  classes.} In the real world, the interfering sounds  and noise often do not belong to our 20 target sound classes. To create a neural network that can generalize to interference from these sounds, we need a diverse set of interfering sound classes in our dataset.\footnote{Note that the target sound classes can also be interfering with each other.} However, this poses several challenges. Firstly, these sounds can come from a very large variety of sources, making it infeasible to exhaustively enumerate all of  them. Secondly, since we want to use them as interfering signals, we must ensure that these sound classes do not overlap with our set of target classes. To overcome these constraints, we use the AudioSet hierarchical structure and our set of 20 target classes to generate a large set of 141 other sound  classes. Specifically, we can define this set as the  nodes that are neither a more specific nor a more general instance of any target (or known) class, according to the AudioSet hierarchy. In other words, by considering the AudioSet ontology as a directed acyclic graph with edges from each sound class node towards its child nodes, we define  unknown sound classes as the set of AudioSet nodes that are disconnected from all target sound class nodes.
\aptLtoX[graphic=no,type=html]{\end{itemize}}{\squishend}

\subsubsection{Audio dataset curation.} Given the sets of the target and other sound classes, we next obtain  labeled audio recordings for each of the  sound classes. The challenge is that, we cannot rely  on only a single general-purpose audio-tagging datasets as was done in prior single-channel work~\cite{waveformer}, This is because  such datasets do not contain audio samples of all 20 target sound classes, and may contain a limited number of audio samples from the other sound classes. So, we combined audio samples from four different datasets: FSD50K~\cite{fsd50k} (general-purpose), ESC-50~\cite{esc50} (environmental sounds), MUSDB18~\cite{musdb18} (music and vocals) and noise files for the DISCO~\cite{disco} dataset (noise sounds).  Since each dataset uses different class names, we standardized the class labels into the AudioSet labels by mapping every class in each dataset to the semantically closest label in the AudioSet ontology, if any. For FSD50K and MUSDB18, we performed additional dataset-specific pre-processing procedures. Specifically, since our goal is to create binaural mixtures of individual sources from multiple directions, we excluded audio samples from FSD50K that were already mixtures of multiple distinct sound sources. For MUSDB18, we extract and split audio into vocal and instrumental streams and assign them the AudioSet labels ``Singing''  and ``Melody,''  respectively. 

We divide the resulting audio samples into 15 second segments and discard all silent ones. We split each dataset into mutually exclusive training, testing, and validation sets and then combined them into our final dataset. For both the FSD50K and MUSDB18, we sample the training and validation audio files from the development split (90-10 split), and the testing samples from the evaluation split. For the ESC-50 dataset, we use the first three folds for training, the fourth fold for validation and the fifth for testing. For the DISCO noise dataset, the audio samples for each sound class is split into train, test and validation sets (60-33-7) before combining with the rest of the datasets. The final combined dataset consists of 20 target classes, distributed as shown in Table~\ref{tab:table1} and 141 other sound classes.

\subsubsection{Binaural data synthesis.} The procedure above describes how we sample single channel sound classes from various audio datasets. However, our goal is to create binaural mixtures that (1) are representative of spatial sounds perceived by a diverse set of listeners, and (2) capture the idiosyncrasies of real world reverberant environments. For this purpose, we use a pre-existing dataset of 43 human head-related transfer function (HRTF) measurements {(CIPIC~\cite{CIPIC}) to address the first challenge. We also augment this with three datasets of measured (SBSBRIR~\cite{sbsbrir}, RRBRIR~\cite{rrbrir})} and simulated (CATT RIR~\cite{CATT_RIR}) reverberant binaural room impulse responses (BRIRs) to address the second requirement. We split each dataset across rooms and listeners into train, test and validation (70-20-10) sets. We ensure that no BRIR subjects or  rooms are sampled across different sets. For each sample during training, we randomly choose one of the datasets and sample a single room and participant from its training set. Then, to create a binaural mixture with $K$ sources, we independently pick a source direction for each of the $K$ sources, out of all source directions available for this room and this participant subject in the dataset. Note that since the source directions are independently picked, two different sound sources might end up being at the same direction from the wearer. We then obtain a set of $2K$  room impulse responses $h_{1, L}, h_{1, R}, h_{2, L}, \dots, h_{K, R} \in \mathbf{R}^{N}$, where $N$ is the length of the room impulse response. Hence, for a training sample with input audio signals of length $T$ samples long, denoted by $x_1, x_2, \dots, x_n \in \mathbf{R}^{T}$, we can compute the sound received at the left and right ears, $s_L$ and  $s_R$ as,
$s_L = \sum_{k = 1}^{K}x_{k} * h_{k, L}$ and  $s_R = \sum_{k = 1}^{K}x_{k} * h_{k, R}$.  Here `$*$' represents the convolution operation. {The synthesized binaural audio mixtures are sampled at 44.1~kHz. If room impulse response in our HRTF dataset has a different sampling rate, we resample the  signal before and after convolving.}

\subsubsection{Training procedure.} We used the Scaper toolkit~\cite{8170052} to synthesize binaural mixtures dynamically on the fly during training. For training and validation, our binaural mixtures consist of two randomly picked target  classes, each with an 5-15~dB~SNR relative to the background sounds, and 1-2 other classes that each have a 0-5~dB~SNR relative to the background sounds. We also  use background sounds sourced from the TAU Urban Acoustic Scenes 2019 dataset~\cite{Mesaros2018_DCASE} in our mixtures. Each mixture is 6 seconds long. Sounds from the target and other background classes are between 3 to 5 seconds long, while background urban sounds persist for the entire duration of the mixture.   In addition to the mixture, we also synthesize  ground truth signals $y_L$ and $y_R$, respectively at the left and right channels, for each chosen target sound source $t$ as,  $y_L = x_{t} * h_{t, L}$ and  $y_R = x_{t} * h_{t, R}$.

The network is then trained to produce a pair of left and right channel target sound estimates $\hat{y}_L$ and $\hat{y}_R$.  To preserve the spatial cues, such as interaural time differences (ITD) and  interaural level differences (ILD), we use the sample-sensitive and scale-sensitive signal-to-noise ratio (SNR) loss function, applied independently and then average the left and right SNRs to obtain the loss function:
\[SNR(\hat{x}, x) = 10\log\left(\frac{||x||^2}{||x - \hat{x}||^2}\right)\]
\[L = -\left(\frac{1}{2}SNR(\hat{y_L}, y_L) + \frac{1}{2}SNR(\hat{y_R}, y_R)\right).\]

Finally, we train our transformer model  for 80 epochs, with an initial learning rate of 5e-4. After completing 40 epochs, we  halve the learning rate if there is no improvement in the validation SNR for more than five epochs. We emphasize here that the {\it training data do not include any measurements with our binaural hardware} and the results we report in this paper evaluate generalization to our hardware, unseen users and environments.
\section{Results}
We  first describe our  setup for real-world evaluations and then present our binaural  network benchmarks.

\vskip 0.05in\noindent{\bf Hardware prototype.} Our hardware setup includes a pair of SonicPresence SP15C binaural microphones that  are wired to capture high-quality recordings. We use an iPhone 12 to process the recorded data and output the audio through noise-canceling headphones like JBL Live 650BTNC and the NUBWO gaming headsets. We use a lightning-to-aux adapter to connect the headphones to the iPhone over a wire. We also use a USB hub to connect both the microphones and the headphones to the smartphone.

\vskip 0.05in\noindent{{\bf Participants.} We recruiting 9 individuals (3 female, 6 male) across our in-the-wild and spatial cues evaluations. We also invited 22 participants (6 female, 16 male) for our online hearing study. 

\subsection{In-the-wild evaluation} \label{sec:wild}

\begin{figure}[t!]
    \includegraphics[width=.38\textwidth]{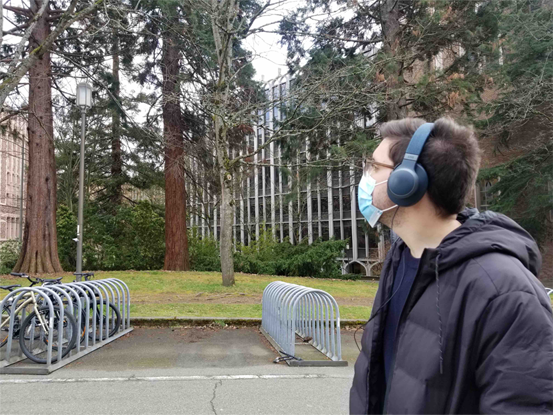}
    \vskip -0.1in
\caption[]{A participant in our in-the-wild evaluation where the target sound was birds  chirping in the presence of  urban environment noises. The participants could move their head freely and the target sound source could also be mobile.}
\label{fig:realworld}
\end{figure}

To evaluate the proposed system in real-life scenarios, we conduct in-the-wild experiments to assess the effectiveness of our system.

{\bf In-the-wild scenarios.}  5 individuals (3 female and 2 male) wore our hardware and collect sounds in the real world. These experiments were conducted in typical application settings: offices, living rooms, streets, rooftops, parks, and restrooms. Since some of the sound classes were relatively less common,  our in-the-wild experiments had a subset of classes which most commonly appeared in our recordings: alarm clock, car horn, door knock, speech, computer typing, hammer, birds chirping, and music. The position and movement of the sound sources were uncontrolled and reflective of real-world scenarios, where the sound sources could be mobile. Furthermore, in all experiments, participants had complete freedom to move their heads, causing the sound source positions relative to the microphones to vary over time (Fig.~\ref{fig:realworld}). Thus, our in-the-wild evaluation captured both mobile wearers as well as mobile sound sources that naturally occured in real-world scenarios (e.g., cars moving or birds that fly).

{\bf Evaluation procedure.} Unlike with our simulated training data, we do not have clean, sample-aligned ground truth signals to objectively compare the binaural outputs of our system with. Hence, we conduct a listening study to compute a mean opinion score (MOS) regarding the sound extraction accuracy.  This metric is crucial to evaluate the perceptual quality of our algorithm for end-users, although it has often been omitted in prior non-speech sound extraction research. 
We invited 22 participants (6 female, 16 male, mean age 34.6) to the online listening study. The study consists of 16 sections. In each section, the participants evaluated the quality of 3 or 4 5.0-8.5 second audio samples. The audio samples played at each section were in-the-wild recordings processed in the following three ways for the same target label: (1) the original recording, (2) the output of our 128-dimensional binaural network,  (3) the output of our 256-dimensional binaural network. For the subset of the evaluations that involved speech as the target sound, we also included an additional fourth audio sample that was obtained by extracting of the interfering class (e.g., door knocks) and then subtracting it from the input recording to estimate the target speech.  

We conducted a pre-screening process to ensure that the participants used suitable binaural headsets. This involved playing two white noise samples, one exclusively from the left channel and one exclusively from the right channel. The participants were instructed to confirm that they heard the sounds only from the correct channels. 
11 of our participants  used headphones, and 11 used earbuds during our online user study. 

We measured the sound extraction quality based on both interference suppression and overall  mean opinion score (MOS), as they are often included in speech enhancement quality assessment:
\begin{enumerate}
    \item \textbf{Noise suppression}: \textit{How INTRUSIVE/NOTICEABLE were the BACKGROUND sounds? 1 - Very intrusive, 2 - Somewhat intrusive, 3 - Noticeable, but not intrusive, 4 - Slightly noticeable, 5 - Not noticeable}
    \item \textbf{Overall MOS}: \textit{If the goal is to focus on the <TARGET> sounds, how was your OVERALL experience? 1 - Bad, 2 - Poor, 3 - Fair, 4 - Good, 5 - Excellent}
\end{enumerate}

{\bf Results.} In Fig.~\ref{fig:analysis}, we present the results of the user evaluations for the interference sound suppression and overall quality improvement of our system for different target sound labels. The results demonstrate the system's capability to significantly reduce unwanted background sounds, as indicated by an increase in the overall noise suppression score from 2.01 (corresponding to \textit{2 - Somewhat intrusive}) to 3.61 (between \textit{3 - Noticeable, but not intrusive} and \textit{4 - Slightly noticeable}) with the 128-dimensional  model, and to 3.84 (slightly worse than \textit{4 - Slightly noticeable}) with the 256-dimensional model. We also observed a similar trend in the overall MOS improvement, with an improvement from 2.63 for the input signal to 3.54 and 3.80 after processing with the 128-dimensional and 256-dimensional models, respectively.  Figs.~\ref{fig:qual_demo}d and \ref{fig:timing} also show  that our network preserves the timing of the target sounds  and can silence out noise outside  the target sound duration.


\begin{figure}[t!]
    \includegraphics[width=.49\textwidth]{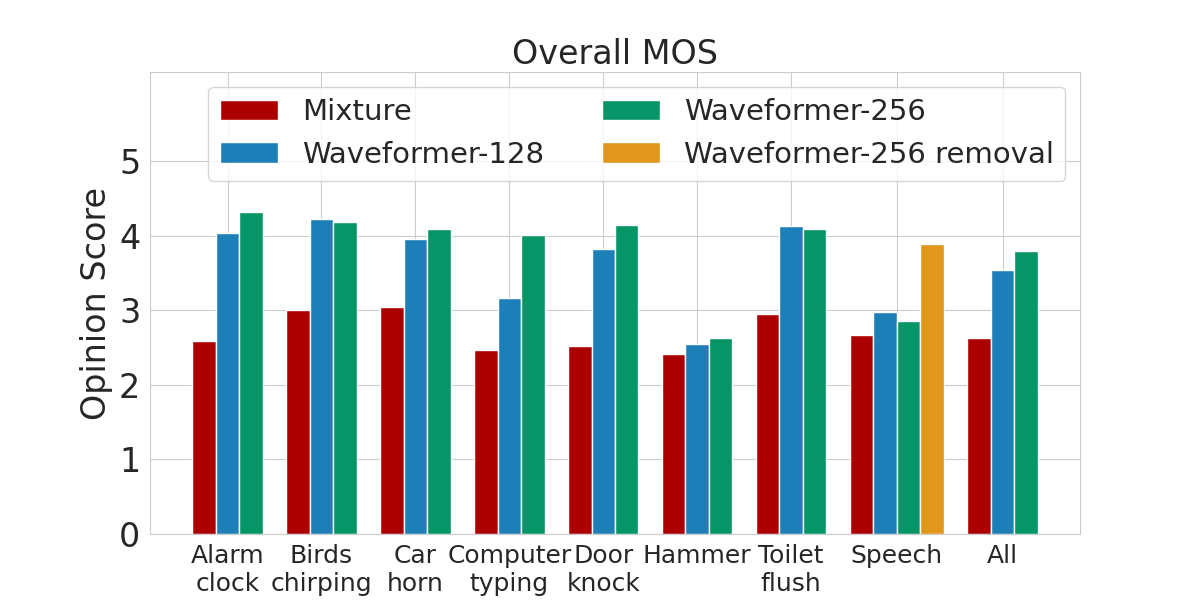}
    \includegraphics[width=.49\textwidth]{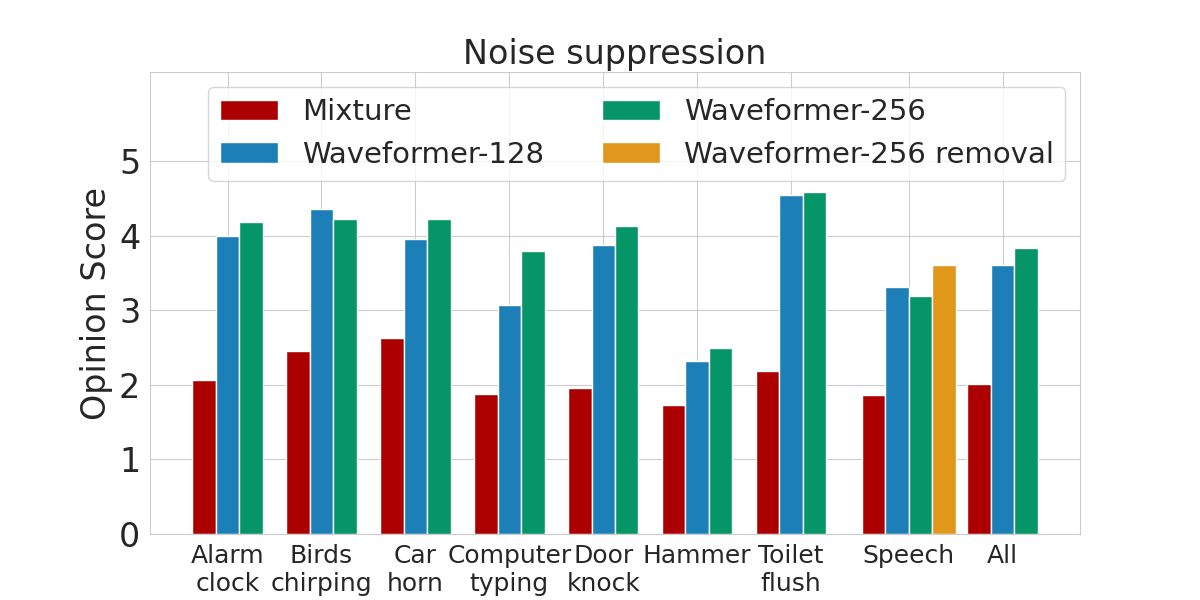}
    \vskip -0.1in
\caption[]{In-the-wild evaluation results for (a) mean opinion score (MOS) and  (b) noise suppression  across various classes that occurred in real-world data collection.}
\vskip -0.15in
\label{fig:analysis}
\end{figure}

\begin{figure}[t!]
\centering
\begin{subfigure}[t]{\linewidth}
    \includegraphics[width=\linewidth]{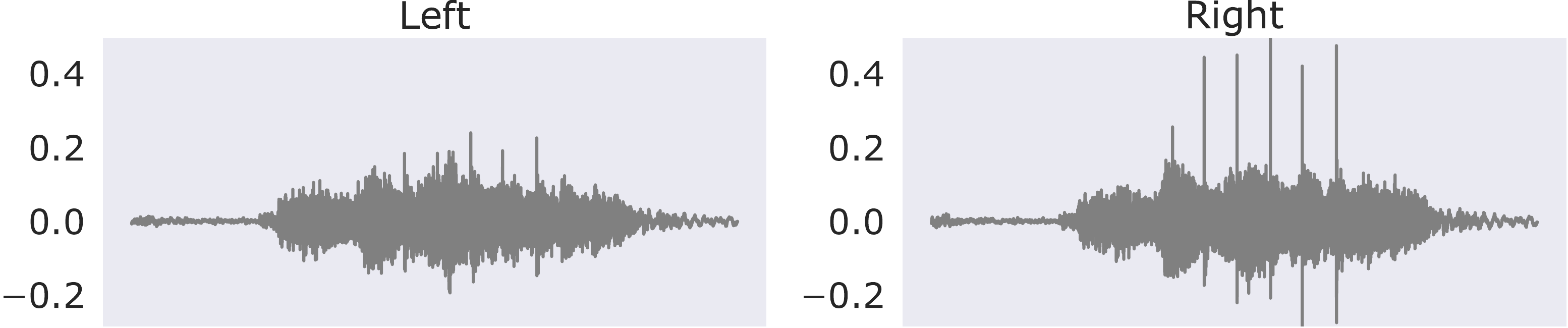}
    \label{fig:top_level}
    \vskip -0.15in
    \caption[]{Binaural input of door knock in the presence of toilet flush.}
\end{subfigure}
\hfill
\centering
\begin{subfigure}[t]{\linewidth}
    \includegraphics[width=\linewidth]{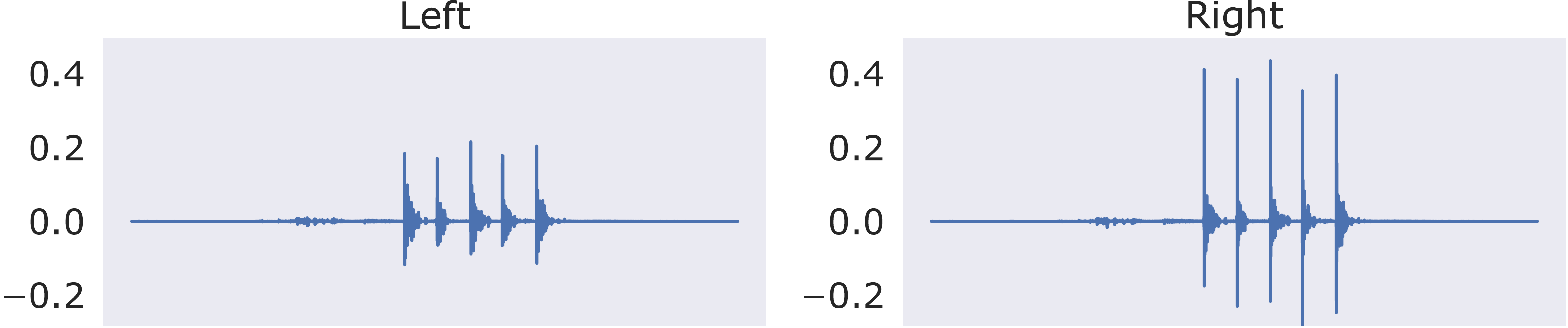}
    \label{fig:enc}
    \vskip -0.15in
    \caption[]{Binaural output with door knock extracted.}
\end{subfigure}
    \vskip -0.1in
    \caption[]{Qualitative result with a real-world recording.}
    \vskip -0.15in
    \label{fig:timing}
\end{figure}

The results also offer interesting insights at a per-class level. In general, the 128-channel model performs only slightly worse than the 256-channel model for almost all classes, except for the ``Computer typing'' class, where the gap in the overall MOS between the two models is almost 0.84 MOS points. This is likely due to a particularly noisy recording taken near a running generator, where the 128-channel model created faint, unpleasant artifacts that were not observed with the 256-channel model. However, both models performed poorly in the ``Hammer'' class, where the target hammer sound was recorded in the presence of  interfering music. Although the network correctly silenced the time segments that did not contain the hammer sounds, there was a noticeable residue from the music when there was a hammer sound, which the listeners found intrusive. Another important finding from the study is the significant improvement obtained by removing interfering signals from the input recording when the target is speech. By removing short-length sounds such as door knocks from the recorded signal instead of extracting the speech directly (see Fig.~\ref{fig:speechsignals}), we were able to increase the overall MOS by 0.91 points. Finally, it's worth noting that these in-the-wild results were obtained from the models trained solely on synthesized data, without any training on data collected from our hardware or for the participants. 

\begin{figure}[t!]
\centering
\begin{subfigure}[t]{\linewidth}
    \includegraphics[width=\linewidth]{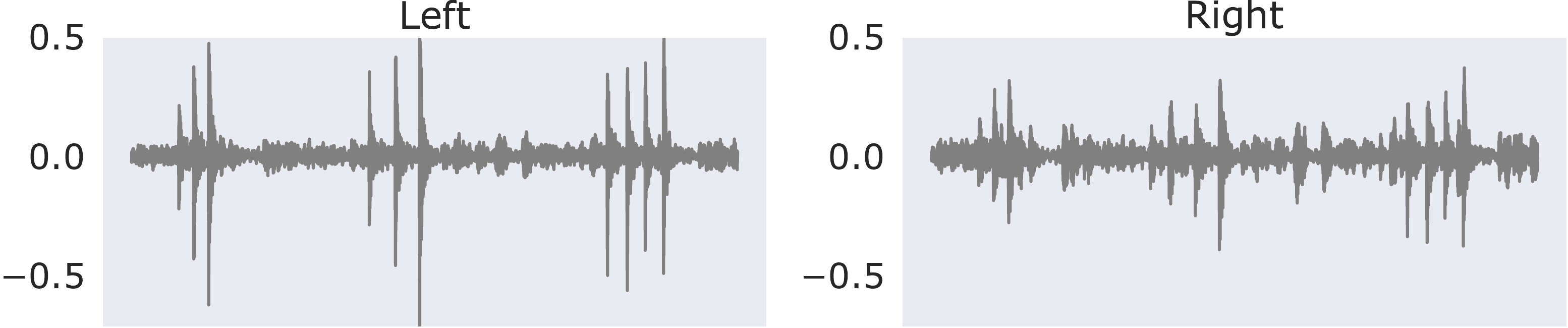}
    \label{fig:top_level}
    \vskip -0.15in
    \caption[]{Binaural input  of speech recorded with  door knock}
\end{subfigure}
\hfill
\centering
\begin{subfigure}[t]{\linewidth}
    \includegraphics[width=\linewidth]{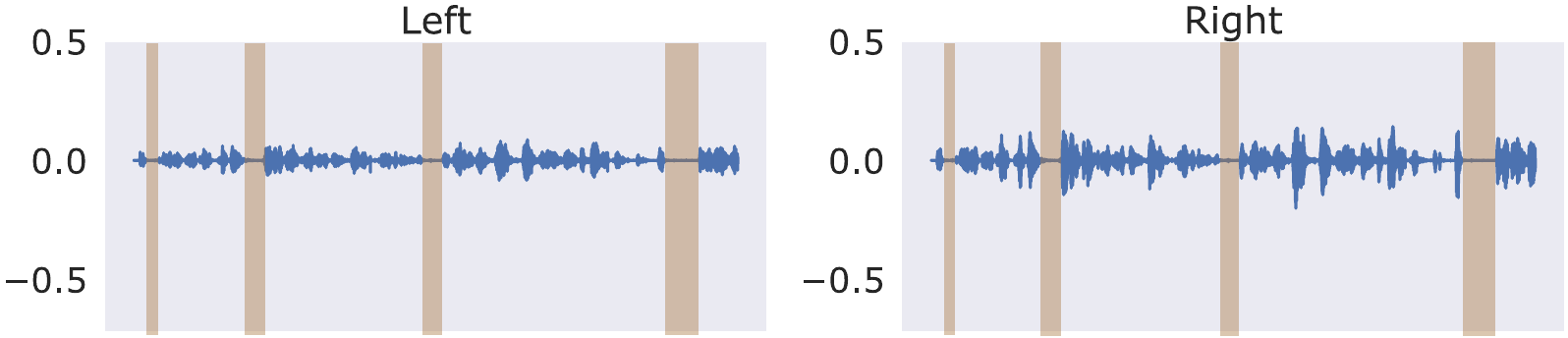}
    \label{fig:enc}
    \vskip -0.15in
    \caption[]{Binaural output with speech extracted}
\end{subfigure}
\hfill
\centering
\begin{subfigure}[t]{\linewidth}
    \includegraphics[width=\linewidth]{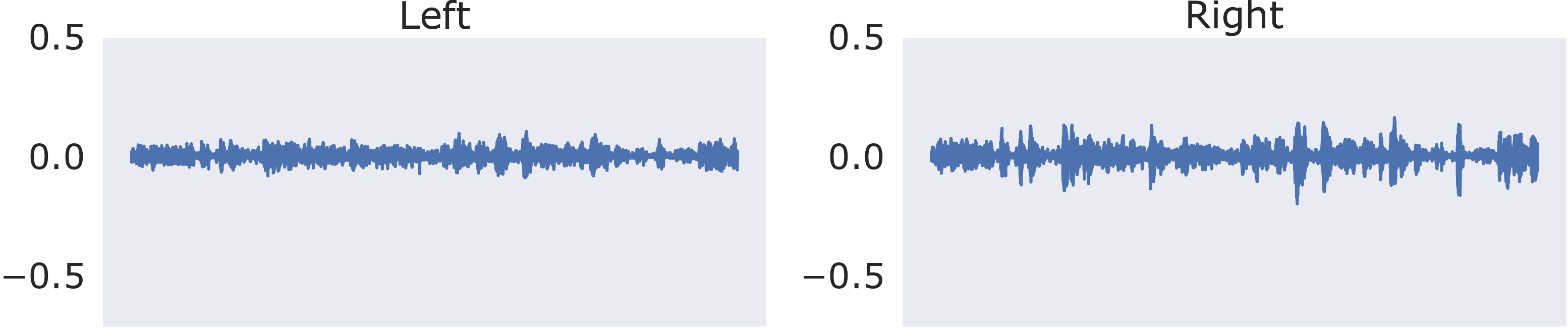}
    \label{fig:enc}
    \vskip -0.15in
    \caption[]{Binaural output with door knock removed}
\end{subfigure}
\vskip -0.1in
    \caption{Extracting speech as a target  here causes  momentary periods of excessive signal attenuation (highlighted in b) as the network tries to remove door knocks and background sounds. However, if we  extract and then subtract door knock sounds, the background noise is still faintly present, and the resulting signal sounds less harsh.}
    \label{fig:speechsignals}
\end{figure}

\subsection{Evaluating {user-perceived} spatial cues}

We present experiments conducted in five ordinary, reverberant rooms to evaluate the ability of our design to preserve or recover user-perceived spatial cues. As with the in-the-wild evaluation,  our training data had  no samples either from our hardware or the tested real-world environments.

\textbf{{Data collection}.} We  collected real-world audio recordings of our target sounds from known directions. To achieve this, {five participants (3 male, 2 female)} were fitted with binaural microphones and seated on a rotating chair positioned at the center of a large, printed semicircular protractor measuring $70\times 36$ inches, as shown in Fig.~\ref{fig:aoa_results}. The protractor was lined at regular 22.5$^\circ$ intervals (nine lines total) for precise rotational measurement. A Sony SRS-XB10 loudspeaker was placed on a fixed tripod at the ${90^\circ}$ line of the protractor to emit different sound signals. To control the angle-of-arrival of the sound signal relative to the listener, the participants were asked to rotate the chair and align themselves with one of the protractor's lines.

The {data was collected} in 9 stages. In each stage, the user is rotated towards a different angle. The first stage starts with the participant facing the $180^\circ$ line. After completion of each stage, the participant rotates $22.5^\circ$ clockwise to the next marked angle. At each stage, the loudspeaker plays four  5-second audio samples: (1) white noise, (2-3) two test samples belonging to the target sound classes, and (4) a test sample belonging to the interfering other  sound classes. Across all stages of  {data collection}, the chosen audio samples comprise exactly 9 test samples from 9 distinct interfering other  sound classes and 6 test samples from 6 distinct target sound classes. Notably, each test sample from the target classes is recorded for 3 different relative angles. 


 
\textbf{Evaluation procedure.} Since our goal is to develop a system that accurately preserves the spatial cues perceived by human listeners, we design a user study to compute the perceived angle-of-arrival for the target binaural sounds output by our system. To this end, based on the collected audio recordings, we first create sound mixtures by sampling two audio clips from the target classes, and 1-2 clips from the interfering  other classes. The mixtures are generated using Scaper. We choose the reference loudness of the background to be -50~LUFS, and we set the SNR of the target class sounds to 15-25~dB and that of the interfering other class sounds to 0-10~dB. We process each mixture by choosing a target class and running the mixture through our  network.

\begin{figure}[t!]
    \centering
    \includegraphics[width=\linewidth]{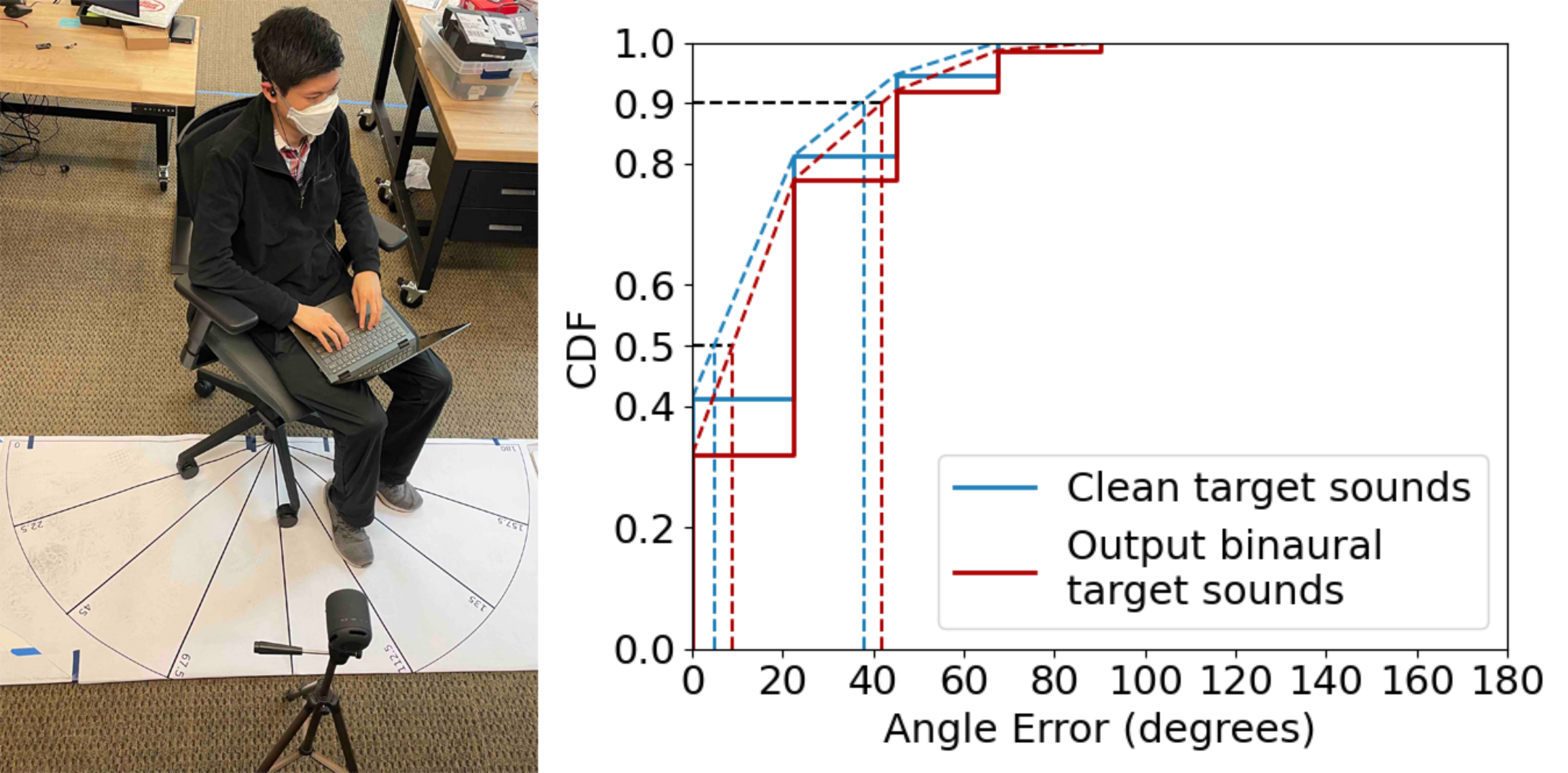}
    \vskip -0.1in
    \caption{Spatial cue evaluation. (left) the evaluation setup, and (right) the CDF of the error between the ground truth source direction and the user-perceived source direction after listening to the isolated clean target sounds as well as network output binaural target sounds. {The dashed lines are interpolated CDFs used to compute the interpolated median and 90th percentile error.}}
    \label{fig:aoa_results}
    \vskip -0.15in
\end{figure}

We {play} the recordings of the individual clean target sounds with no interference, as well as the network output samples estimating these target sounds { from the created mixtures}, to the { same set of} participants via a pair of binaural earphones. {Since the perceived spatial cues rely heavily on anthropometric features, all the sound signals played to a given participant originated from the binaural data obtained from that same participant in the data collection step.} Prior to listening to each sample, participants are informed of the target class they should be localizing. After listening, they are asked to predict the direction of the sound source. To prevent the participants from associating each output sample with its corresponding individually-recorded target sound, the samples are played in a random order. To help the participants establish an orientation reference, we play back the white noise samples for each angle in the increasing order at the start of the evaluation. Additionally, in cases of uncertainty between two specific source angles, the participants are allowed to re-listen to the white noise samples recorded for these angles. {The study lasted around 20 minutes per user.}

\textbf{Results.} We compare the errors between the ground truth source directions and the users' perceived arrival directions obtained for both the clean interference-free target sound recordings as well as the  binaural target sound signals generated by our system for the mixture signal input. Our findings, as illustrated in Fig.~\ref{fig:aoa_results},  {show that the mean angle error slightly increases from $18^\circ$ to $23.25^\circ$. Additionally, we observe that the interpolated 50th and 90th percentile errors also increase marginally from $5^\circ$ to $9^\circ$ and from $38^\circ$ to $42^\circ$, respectively.} This demonstrates that our model preserves the spatial cues of the target sounds in its output and has a negligible impact on how users perceive the source directions. 

\subsection{Integration with noise canceling headsets}

\begin{figure}[t!]
\centering
\begin{subfigure}[t]{\linewidth}
    \includegraphics[width=\linewidth]{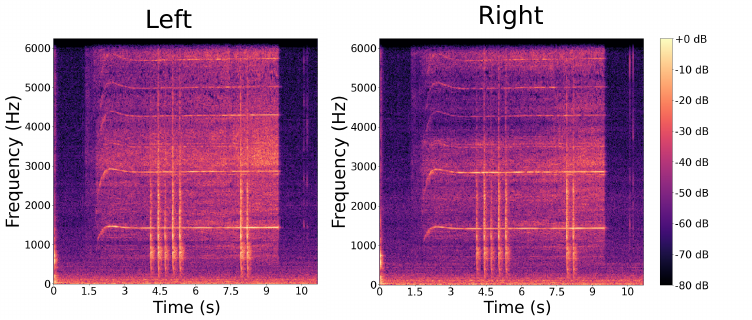}
    \label{fig:spectra_outer}
    \vskip -0.15in
    \caption[]{Sound recorded outside the headphone cups.}
\end{subfigure}
\hfill
\centering
\begin{subfigure}[t]{\linewidth}
    \includegraphics[width=\linewidth]{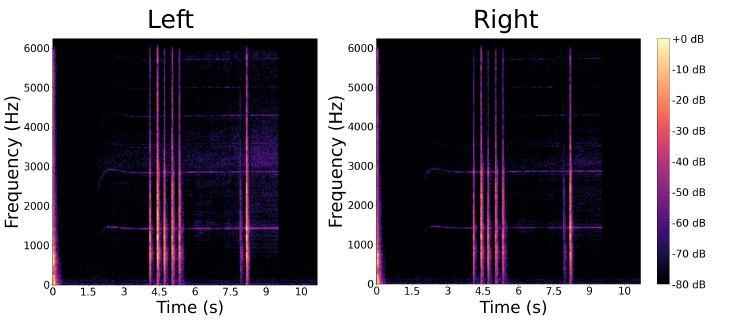}
    \label{fig:spectra_output}
    \vskip -0.15in
    \caption[]{Semantic hearing output played by headphones.}
\end{subfigure}
\hfill
\centering
\begin{subfigure}[t]{\linewidth}
    \includegraphics[width=\linewidth]{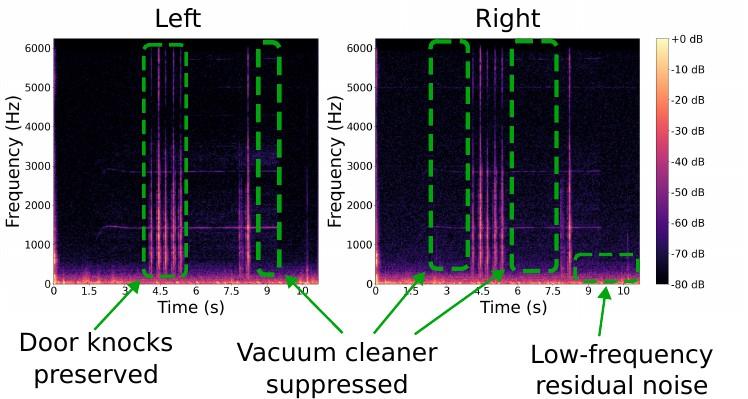}
    \label{fig:spectra_inner}
    \vskip -0.15in
    \caption[]{Sound recorded inside the headphone cups.}
\end{subfigure}
\vskip -0.1in
    \caption[]{Spectrograms of binaural recordings showing results from our end-to-end experiment with a wearable headset. Here, we extract door knock sounds in an environment with a nearby active vacuum cleaner.}
    \vskip -0.15in
    \label{fig:spectra}
\end{figure}

{So far, we have  treated  semantic hearing and active noise cancellation  as two separate systems that function independently. In practice, however, the end-to-end system requires a few additional considerations. Firstly, many active noise cancellation systems rely on a recorded signal inside the ear cup to adaptively silence the noise signals and achieve adaptive noise cancelation. Hence, the audio we play back to perform semantic hearing may influence the noise cancellation algorithm. Secondly, active noise cancellation systems are not perfect, and they may still let some sounds through. To address these concerns, we record data while a user is utilizing our end-to-end system in real time. The user wears a pair of Sony WH-1000XM4 headphones with active noise cancellation enabled. In addition to the outer microphones used to capture external sounds to process, they also wear binaural microphones inside the earcups to record the sound produced by the active noise cancellation and semantic hearing systems together, i.e. as heard by the user. The user chooses to listen to the sound of door knocks as a vacuum cleaner is turned on nearby. For this experiment alone, we run our semantic hearing algorithm on the audio recorded from the outer microphones on a laptop with an Intel Core i5 CPU. The processed audio is played back through the headphones.}

Fig.~\ref{fig:spectra}(a)-(c) shows the spectrograms for three binaural signals: the signal recorded at the outer microphone, the signal played through the headphones, and the signal recorded inside the earcups. We demonstrate that while the recordings from the inside earcups are slightly noisier, we clearly see that the system can  suppress the unwanted sounds (vacuum cleaner), while preserving the target sounds (door knocks). This demonstrates the feasibility that   such a system can coexist with active noise cancellation systems. We note that to mitigate  residual noises, the semantic hearing subsystem may have to integrate the residual audio from noise cancelling headphones to adapt the playback signal to the residual noise as well. However, this comes with stricter latency requirements and thus we leave it for future work.

\subsection{Benchmarking the neural network}\label{sec:benchmark}
In-the-wild evaluation with human evaluators is closest to real-world use. It is however hard to objectively compare different models due to the lack of ground-truth signals, as well as due to the challenges in obtaining a large amount of test data necessary for the statistical significance of smaller performance gaps. To address these practical limitations, we also evaluate our model on an extensive reverberant binaural testset comprising 10000 mixture and ground-truth pairs. We synthesized the benchmarking dataset to mimic real-world situations following the  approach  in \xref{training}. 

\begin{table*}[!ht]
    \centering
    \caption[]{Performance and efficiency comparison of different binaural target sound extraction frameworks and mask estimation architectures on a large test dataset across 20 target classes generated using the approach described in \xref{training}.}
    \begin{tabular}{lcccccc}
    \toprule
    Binaural framework & Mask estimator & Params (M) & SI-SNRi (dB) & $\Delta$ITD ($\mu s$) & $\Delta$ILD (dB) & Runtime (ms) \\
    \midrule
     Dual-ch &  Ours ($D=128$) & 0.52 & 7.17 & 87.77 & 0.88 & 6.56 \\
             &  Ours ($D=256$) & 1.74 & 7.41 & 85.16 & 0.87 & 12.54 \\
    \midrule
     Parallel & Ours ($D=128$) & 0.86 & 7.24 & 81.72 & 1.08 & 13.35 \\
              & Conv-TasNet & 2.33 & 4.43 & 670.05 & - & 15.58 \\
    \midrule
     Single-ch & Ours ($D=256$) & 1.68 & 7.43 & 79.70 & 1.32 & 22.19 \\
               & {Vanilla Waveformer ($D=256$) \cite{waveformer}} & 1.69 & 7.37 & 85.33 & 1.27 & 25.85 \\
    \bottomrule
    \end{tabular}
    \label{benchmark}
\end{table*}

To evaluate the performance of our binaural extraction model, as shown in  Table \ref{benchmark}, we compare the following three binaural target sound extraction frameworks.  
\aptLtoX[graphic=no,type=html]{\begin{itemize}}{\squishlist}
    \item \textbf{Dual-ch.} This is the dual-channel architecture we proposed in \xref{framework} for efficient binaural target sound extraction. In this framework, the binaural signal is converted into a combined latent space representation before the mask estimation. Since both left and right channels are combined into a common representation, a single instance of the mask estimation network is used for estimating the mask corresponding to the target sound. We consider our mask estimation architecture with both $D = 128$ and $D = 256$.
    \item \textbf{Parallel.} This is the binaural framework proposed in \cite{han2020realtime} that implements parallel processing of the left and right channels, along with some cross-communication between channels. The binaural framework in \cite{han2020realtime} is originally proposed for binaural speech separation. We  implemented this framework for both our mask estimation network with $D = 128$ and Conv-TasNet \cite{luo2019conv}. We include Conv-TasNet as it is one of the most widely used signal enhancement model architectures. We choose a configuration of Conv-TasNet that resulted in similar runtime to that of our model and trained both models with our training dataset. 
    \item \textbf{Single-ch.} In addition to the above two binaural extraction frameworks, we also evaluate and compare the performance with a single-channel extraction baseline. Since the target sound extraction models we consider are sample-aligned, models trained with monaural inputs and outputs can be independently applied to the left and right channels. Similar to the \emph{Parallel} case, this also involves two instances of the mask estimation network. However, by contrast, the model parameters applied to the left and right channels are the same and there is no cross-communication between the channels. We implement the best configuration of our model ($D = 256$) so that this serves as a strong baseline.
\aptLtoX[graphic=no,type=html]{\end{itemize}}{\squishend}

For each model, we compare the performance in terms of the signal quality, the accuracy in spatial cues, and the  on-device  runtime requirement. We measure the signal quality using the scale-invariant signal-to-noise-ratio \cite{https://doi.org/10.48550/arxiv.1811.02508} improvement (SI-SNRi) of the output compared to that of the mixture, computed with respect to the ground-truth. The SI-SNRi results are averaged over the entire testset, across the left and right channels. Following \cite{han2020realtime}, the spatial cue accuracy is measured using the difference in the interaural time differences (ITDs) and interaural level differences (ILDs) between the output binaural signal and the ground-truth binaural signal, denoted as $\Delta$ITD and $\Delta$ILD. {We compute  ITD using cross-correlation, limiting  them to $\pm$1~ms, as was done in \cite{itd_matlab}.} The model runtimes  are measured on iPhone 11, by converting them to ONNX format \cite{bai2019} and then executing them using ONNX Runtime for iOS. The runtimes are measured for computing a 10 ms output chunk averaged over 100 runs. Therefore, the runtime must be less than 10 ms for deployment, which our dual-channel model with $D=128$ meets.

In our experiments, we observed that the causal Conv-TasNet converges to the local minima of generating a constant zero signal when trained only with the SNR loss. This phenomenon is also observed in \cite{waveformer}, which suggested training Conv-TasNet with 90\% SNR + 10\% SI-SNR loss. The likely cause for this is, unlike the speech datasets that  Conv-TasNet is originally designed for, sound datasets have a significant amount of silence, causing the Conv-TasNet optimization process to converge to generating a zero signal. On the other hand, using a  loss of 90\% SNR + 10\% SI-SNR in the binaural case, caused one of the channels to output a very low-amplitude signal relative to the other channel as 
SI-SNR is insensitive to the signal gains. We confirmed that the signal is spectrally meaningful even though the magnitude is wrong. As a result, only SI-SNRi and $\Delta$ITD results are meaningful for the Conv-TasNet model. $\Delta$ILD computation resulted in infinity, so we omit it in our table. 

In Table.~\ref{benchmark}, we observe that the dual-channel framework is competitive with the parallel and single-channel frameworks in terms of  SI-SNRi, while outperforming in $\Delta$ILD. 
With regard to $\Delta$ITD, it resulted in a very marginal increase. 
These results intuitively make sense because the dual-channel framework has a sample-aligned common representation for both left and right channels. As a result, it can maintain the relative amplitudes of the left and right channels. On the other hand, the parallel and single-channel frameworks have separate branches that independently process different channels,  facilitating maintaining the sample alignment with the respective channels. This phenomenon is more notable for the single-channel framework, where the SI-SNRi and $\Delta$ITD are promising but the $\Delta$ILD is poor, as there is no cross-communication between the left and right channel processings. We note that our dual-channel framework requires only a little more than 50\%  of the runtime required by their parallel or single-channel counterparts, making it a good practical choice for our  semantic hearing system. {Finally, we note that  our dual-channel framework uses 240 MFLOPS, while vanilla  Waveformer uses 357 MFLOPS across the two microphones.}

{For our causal model, the receptive field is exclusively the past audio. Hence, it has no effect on the algorithmic latency. The algorithmic latency of our model is the sum of chunk size, $KL$, and the lookahead of the input convolution, $L$, where $L$ is the stride of the input convolution (\xref{streaming}). Table \ref{benchmark} uses stride $L = 32$ samples and $K = 13$, resulting in a chunk size of $KL = 416$ samples and lookahead $L = 32$ samples. This is equivalent to 9.4 ms and 0.7 ms, respectively. Table \ref{latency} shows the performance of our binaural model with various chunk sizes to understand the effect of algorithmic latency on performance. The results show that our model achieves reasonable performance with an algorithmic latency as low as 1.4 ms. Thus with ASIC implementations, such as those in hearing aids, we could envision ultra-low-latency semantic hearing systems.}

{During our in-the-wild evaluations, users freely moved their heads and encountered mobile sources (eg. sirens). The model also performed robustly without glitches during evaluations by human testers. The model adapted quickly to relative motion because it outputs small chunks (<10ms) while updating its internal state. The model can also utilize spatial positions in the trajectory that have better level differences between L and R channels. In addition to that qualitative evaluation, Table \ref{motion_results} provides a quantitative comparison of the performance for different amounts of relative angular motion between the listener and sound sources. For this comparison, we use the dual-ch model with $D = 256$ dimensions. We simulate motion using Steam Audio SDK \cite{steamaudio-sdk}, which simulates binaural motion given an HRTF file in the SimpleFreeFieldHRIR format \cite{simplefreefield-hrir}. We performed controlled experiments with different angular velocities in both anechoic and reverberant environments, with sources moving from a random position on an arc with the given angular velocity. We used the CIPIC \cite{CIPIC} HRTF dataset for anechoic simulations and RRBRIR \cite{rrbrir} BRIR dataset for reverberant simulations as they provide impulse responses in the SimpleFreeFieldHRIR format. We synthesize the binaural audio in frames of 1024 samples by convolving with an interpolated impulse response using bilinear interpolation at every frame. Since the ILD and ITD are now time-varying, we compute the $\Delta$ILD and $\Delta$ITD in chunks of 250~ms, discarding any chunks where the clean signal is silent on both channels, and take the mean across the remaining chunks.  We observed that in the presence of motion, SI-SNRi and $\Delta$ILD are marginally better as the model is able to better leverage the level differences between L and R at different relative angular positions while achieving lower $\Delta$ITD in the anechoic case and slightly higher $\Delta$ITD in reverberant scenarios.}}

\begin{table}[!t]
    \centering
    \caption[]{{Effect of algorithmic latency on the performance. *Proposed system with  end-to-end latency $\sim$20 ms.}}
    \vskip -0.1in
    \begin{tabular}{ccc}
    \toprule
    Chunk size (samples) & Algorithmic latency (ms) & SI-SNRi (dB) \\
    \midrule
    32 & 1.4 & 6.59 \\
    128 & 3.6 & 6.83 \\
    256 & 6.5 & 7.18 \\
    416* & 10.1 & 7.42 \\
    \bottomrule
    \end{tabular}
    \vskip -0.1in
    \label{latency}
\end{table}

\begin{table}[!t]
    \centering
    \caption[]{{Comparison of performance in the presence of relative angular motion between listener and sound sources. Dual-ch model with $D = 256$ is used for this evaluation.}}
    \vskip -0.1in
    \begin{tabular}{ccccc}
    \toprule
    Angular & Reverb. & SI-SNRi (dB) & {$\Delta$ITD ($\mu$s)} & {$\Delta$ILD (dB)} \\
    velocity ($^\circ$/s) & & & & \\
    \midrule
    30 & No & 7.95 & 34.26 & 0.58 \\
       & Yes & 7.88 & 103.45 & 0.43 \\
    \midrule
    60 & No & 7.91 & 49.49 & 0.57 \\
       & Yes & 7.98 & 98.23 & 0.49 \\
    \midrule
    90 & No & 7.87 & 58.67 & 0.54 \\
       & Yes & 8.00 & 99.83 & 0.43 \\
    \bottomrule
    \end{tabular}
    \vskip -0.1in
    \label{motion_results}
\end{table}

\subsection{Proof-of-concept user interface}
\begin{figure}
    \centering
    \includegraphics[width=\linewidth]{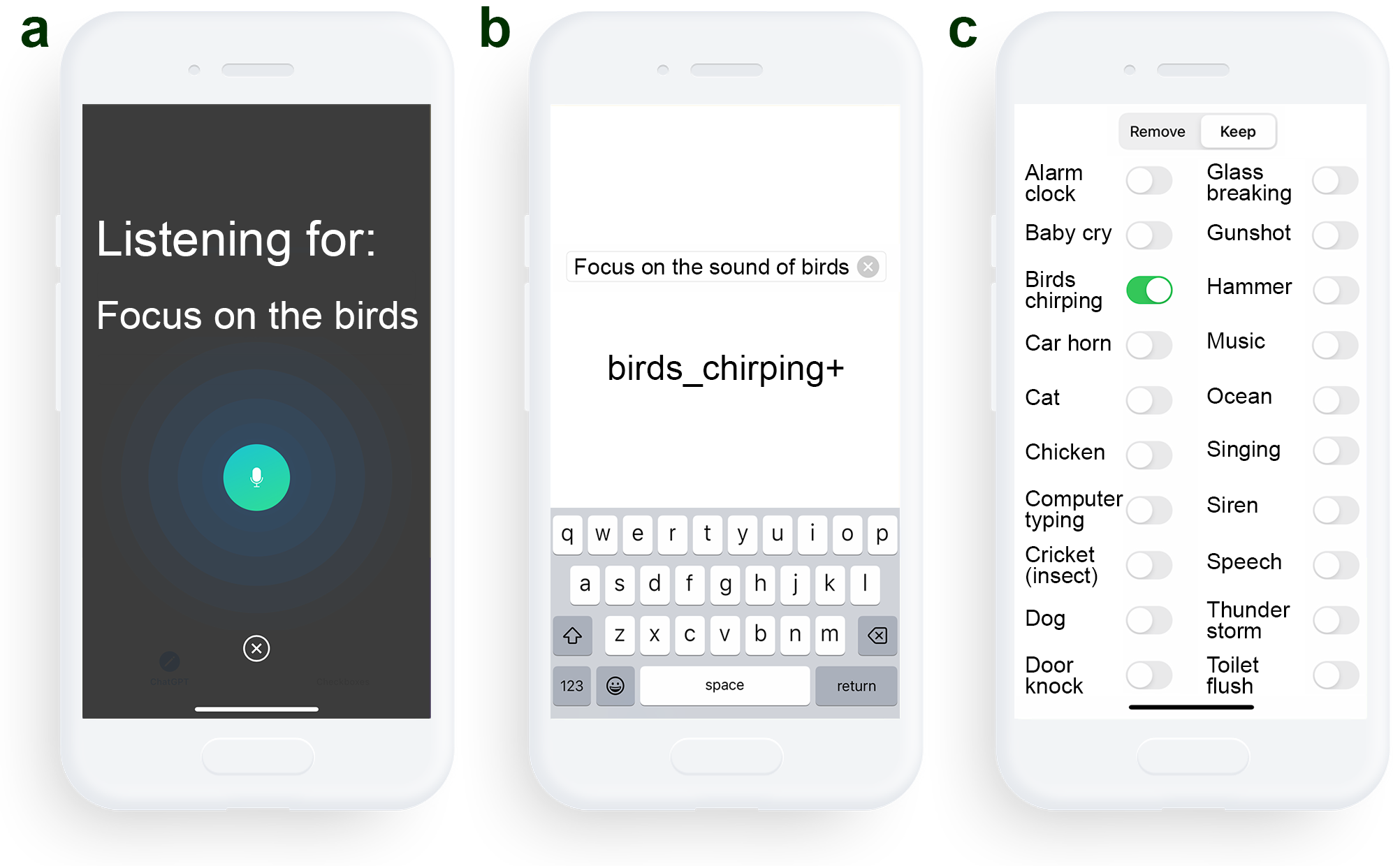}
    \vskip -0.15in
    \caption[]{{User interface designs for target sound selection on a smartphone. Each design uses a different input method to capture the user’s intent: (left to right) speech, text and toggle switches. The first two interfaces use ChatGPT API to convert natural language to class inputs for  our system.}}
\vskip -0.15in
    \label{fig:uis}
\end{figure}

{Finally, a natural question is: how does the user pick between classes? To answer this question, we prototyped an iOS app with three different user interfaces for sound selection: Speech, Text and Toggle switch grid of sounds (Fig.~\ref{fig:uis}), and  evaluated their accuracy, speed, and ease of use.}

{For the speech and text interfaces, our goal was to investigate if the ChatGPT API for phones~\cite{chatgpt_api} could be used to convert natural language (\textit{I want to listen to ambulance sounds}) into known sound class inputs to our system (\textit{siren+}). To do this, we initialized ChatGPT using the prompt: \textit{Here is a list of sound classes: [`alarm\_clock', `baby\_cry', [...] I will provide you a sentence that involves keeping or removing one of these sound classes. I want you to output the sound class from the list that most closely matches (semantically) the sounds in the sentence. If there are no classes that are sufficiently close, output `na'. Please do not output any other characters. If you find a close class and if the sentence involves keeping sound from this label, append a `+' to your output, otherwise, append `-'. For example, if I say `mute cat' you should say `cat-'. If I say `mute cow' you should say `na'.}}

{Ten participants were presented each of them with ten scenes of the following form: \textit{Your infant's cries break the silence. Your phone plays a melody. The rustle of the wind is audible.} We asked participants to select a single sound event to add or remove, and convey their intent to the app with the three user interfaces (UIs). To evaluate the accuracy of each interface, we compared the sound event selected through each UI with our best interpretation of what the user said. Agreement rates were 92\% for Speech and Text, and 93\% for Toggle switch. For Speech and Text, disagreement was due to confusion by ChatGPT (e.g. \textit{The toilet is too loud} mapped to \textit{toilet\_flush+}), or when ChatGPT would map selected sounds not in the dataset to a similar sound in the dataset (e.g. \textit{wind} and \textit{fountain sounds} were mapped to \textit{ocean}). For the Toggle switch, disagreement occurred when the intended sound class could not be found.} 

{The mean time taken to convey intent was shortest for Speech ($5.5 \pm 1.0$s), then Toggle ($6.3 \pm 3.3$s) and longest for Text ($8.3 \pm 3.7$s). Preference ratings (1=very unlikely, 5=very likely) were highest for Speech ($4.0 \pm 1.1$), then Text ($2.9 \pm 1.2$), and lowest for Toggle ($2.7 \pm 1.4$). These findings suggest that from a user interface perspective, Speech would be a practical interface choice and would scale better than the Toggle interface as the number of supported classes increases. One participant noted that they would prefer the Text interface when using the system in a public setting even if it took a longer time to input their intent.}


\section{Limitations and Discussion}

As shown in Table.~\ref{tab:table1},  we have an imbalance in the number of examples across classes. For instance, the ``speech'' class has 494 training examples, while ``car horn'' has only 60 training examples. Collecting more examples across all classes can potentially improve the performance. Finally, some classes may be inherently harder to separate. For example, music and human speech share many characteristics, including vocal sounds and harmonicity. Thus, despite having a larger number of training examples, it is difficult for our model to perform tasks such as separating the speech of the person around the wearer in the presence of background music that also has vocals. Similarly, it can be challenging to separate music from other classes like alarm clock sounds or bird chirping. Additionally, our training methodology does not utilize any real-world data with our hardware. Nevertheless, our real-world testing results demonstrate the generalization capabilities to our hearable hardware as well as unseen real-world environments. However, it is still possible that collecting training data in the real-world scenarios as well as with actual hardware can help improve the system performance.

Another limitation is the form factor of the hearable hardware we used in our evaluations where we used binaural earphones in addition to a noise canceling headset. {The form factor could be simplified if we used a single device for recording and playback. Currently, there are commercial noise cancelling headsets that provide user access to the microphone data, such as the Sennheiser AMBEO Smart Headset, which we found after our evaluations. Our system implemented on such a device would have fewer wires and would directly connect to the smartphone at a single point, without the need for an additional pair of binaural earphones.} 



Binaural target sound extraction can also be used to   subtract the target sounds and play the residual sounds into the ear.  Fig.~\ref{fig:speechsignals}  shows the results for subtracting a target sound (e.g., computer typing or hammer) to focus on the human speech. This can be beneficial when the user knows the specific type of environmental noise that they feel annoying (e.g., computer typing in an office room) as this approach would remove only the specified noise and thus allow the user to focus on the speech and the other sounds in the environment. 

For proof-of-concept demonstration, we have implemented our neural network on a connected smartphone. While wired headsets connected to the smartphones are an important use case for practical applications and can benefit from our implementation, extending our system to wireless headsets requires integrating computation with the headset hardware itself. This is likely feasible given the ultra-low power multicore embedded GPUs that are being designed for wearable devices~\cite{gpuwear}. Further, with recent developments in custom silicon for on-chip deep learning for speech and natural language processing~\cite{silicon}, it is likely that commercial hearable devices for semantic hearing would use such custom silicon to reduce both the power consumption of the wearable device and  the end-to-end latency.


\section{Conclusion}

This paper takes an important first step towards realizing real-time programming of acoustic scenes on binaural hearable devices using the semantic description  of sounds.      At its core are two key technical contributions: 1) the first binaural target sound extraction neural network. Our network can run in real-time, using 10 ms or less of audio blocks, while preserving the spatial information, and 2) a training methodology that allows our system to generalize to unseen real-world environments. In-the-wild experiments with participants show that our proof-of-concept hardware-software system can preserve the directions of the target sounds and separate these sounds in real-time from both the background noise and other sounds in the environment.


\bibliographystyle{ACM-Reference-Format}
\bibliography{ref}


\begin{thebibliography}{71}


\ifx \showCODEN    \undefined \def \showCODEN     #1{\unskip}     \fi
\ifx \showDOI      \undefined \def \showDOI       #1{#1}\fi
\ifx \showISBNx    \undefined \def \showISBNx     #1{\unskip}     \fi
\ifx \showISBNxiii \undefined \def \showISBNxiii  #1{\unskip}     \fi
\ifx \showISSN     \undefined \def \showISSN      #1{\unskip}     \fi
\ifx \showLCCN     \undefined \def \showLCCN      #1{\unskip}     \fi
\ifx \shownote     \undefined \def \shownote      #1{#1}          \fi
\ifx \showarticletitle \undefined \def \showarticletitle #1{#1}   \fi
\ifx \showURL      \undefined \def \showURL       {\relax}        \fi
\providecommand\bibfield[2]{#2}
\providecommand\bibinfo[2]{#2}
\providecommand\natexlab[1]{#1}
\providecommand\showeprint[2][]{arXiv:#2}

\bibitem[\protect\citeauthoryear{??}{air}{2023}]%
        {airpods}
 \bibinfo{year}{2023}\natexlab{}.
\newblock \bibinfo{title}{Apple {A}ir{P}ods}.
\newblock \bibinfo{howpublished}{\url{https://www.apple.com/airpods/}}.
  (\bibinfo{year}{2023}).
\newblock


\bibitem[\protect\citeauthoryear{??}{ios}{2023}]%
        {ioslatency}
 \bibinfo{year}{2023}\natexlab{}.
\newblock \bibinfo{title}{Audio Latency Meter for i{OS}}.
\newblock \bibinfo{howpublished}{\url{https://onyx3.com/LatencyMeter/}}.
  (\bibinfo{year}{2023}).
\newblock


\bibitem[\protect\citeauthoryear{??}{ada}{2023}]%
        {adaptivetransparency}
 \bibinfo{year}{2023}\natexlab{}.
\newblock \bibinfo{title}{Customize Transparency mode for {A}ir{P}ods {P}ro}.
\newblock
  \bibinfo{howpublished}{\url{https://support.apple.com/guide/airpods/customize-transparency-mode-dev966f5f818/web}}.
    (\bibinfo{year}{2023}).
\newblock


\bibitem[\protect\citeauthoryear{??}{cha}{2023}]%
        {chatgpt_api}
 \bibinfo{year}{2023}\natexlab{}.
\newblock \bibinfo{title}{GPT models}.
\newblock
  \bibinfo{howpublished}{\url{https://platform.openai.com/docs/guides/gpt}}.
  (\bibinfo{year}{2023}).
\newblock


\bibitem[\protect\citeauthoryear{??}{gpu}{2023}]%
        {gpuwear}
 \bibinfo{year}{2023}\natexlab{}.
\newblock \bibinfo{title}{{GPU-WEAR}, Ultra-low power heterogeneous Graphics
  Processing Units for Wearable/IoT devices}.
\newblock
  \bibinfo{howpublished}{\url{https://cordis.europa.eu/project/id/717850}}.
  (\bibinfo{year}{2023}).
\newblock


\bibitem[\protect\citeauthoryear{??}{sim}{2023}]%
        {simplefreefield-hrir}
 \bibinfo{year}{2023}\natexlab{}.
\newblock \bibinfo{title}{SimpleFreeFieldHRIR}.
\newblock
  \bibinfo{howpublished}{\url{https://www.sofaconventions.org/mediawiki/index.php}}.
    (\bibinfo{year}{2023}).
\newblock


\bibitem[\protect\citeauthoryear{??}{ste}{2023}]%
        {steamaudio-sdk}
 \bibinfo{year}{2023}\natexlab{}.
\newblock \bibinfo{title}{Steam Audio SDK}.
\newblock
  \bibinfo{howpublished}{\url{https://valvesoftware.github.io/steam-audio/}}.
  (\bibinfo{year}{2023}).
\newblock


\bibitem[\protect\citeauthoryear{Algazi, Duda, Thompson, and Avendano}{Algazi
  et~al\mbox{.}}{2001}]%
        {CIPIC}
\bibfield{author}{\bibinfo{person}{V.R. Algazi}, \bibinfo{person}{R.O. Duda},
  \bibinfo{person}{D.M. Thompson}, {and} \bibinfo{person}{C. Avendano}.}
  \bibinfo{year}{2001}\natexlab{}.
\newblock \showarticletitle{The CIPIC HRTF database}. In
  \bibinfo{booktitle}{{\em Proceedings of the 2001 IEEE Workshop on the
  Applications of Signal Processing to Audio and Acoustics (Cat.
  No.01TH8575)}}. \bibinfo{pages}{99--102}.
\newblock
\showDOI{%
\url{https://doi.org/10.1109/ASPAA.2001.969552}}


\bibitem[\protect\citeauthoryear{Bai, Lu, Zhang, et~al\mbox{.}}{Bai
  et~al\mbox{.}}{2019}]%
        {bai2019}
\bibfield{author}{\bibinfo{person}{Junjie Bai}, \bibinfo{person}{Fang Lu},
  \bibinfo{person}{Ke Zhang}, {et~al\mbox{.}}} \bibinfo{year}{2019}\natexlab{}.
\newblock \bibinfo{title}{{ONNX}: Open Neural Network Exchange}.
\newblock \bibinfo{howpublished}{\url{https://github.com/onnx/onnx}}.
  (\bibinfo{year}{2019}).
\newblock


\bibitem[\protect\citeauthoryear{Brayda, Traverso, Giuliani, Diotalevi,
  Repetto, Sansalone, Trucco, and Sandini}{Brayda et~al\mbox{.}}{2015}]%
        {brayda2015spatially}
\bibfield{author}{\bibinfo{person}{Luca Brayda}, \bibinfo{person}{Federico
  Traverso}, \bibinfo{person}{Luca Giuliani}, \bibinfo{person}{Francesco
  Diotalevi}, \bibinfo{person}{Stefania Repetto}, \bibinfo{person}{Sara
  Sansalone}, \bibinfo{person}{Andrea Trucco}, {and} \bibinfo{person}{Giulio
  Sandini}.} \bibinfo{year}{2015}\natexlab{}.
\newblock \showarticletitle{Spatially selective binaural hearing aids}. In
  \bibinfo{booktitle}{{\em Adjunct Proceedings of IMWUT/ISWC}}.
\newblock


\bibitem[\protect\citeauthoryear{Bui, Pham, Barnitz, Zou, Nguyen, Truong, Kim,
  Farrow, Nguyen, Xiao, Deterding, Dinh, and Vu}{Bui et~al\mbox{.}}{2021}]%
        {tam1}
\bibfield{author}{\bibinfo{person}{Nam Bui}, \bibinfo{person}{Nhat Pham},
  \bibinfo{person}{Jessica~Jacqueline Barnitz}, \bibinfo{person}{Zhanan Zou},
  \bibinfo{person}{Phuc Nguyen}, \bibinfo{person}{Hoang Truong},
  \bibinfo{person}{Taeho Kim}, \bibinfo{person}{Nicholas Farrow},
  \bibinfo{person}{Anh Nguyen}, \bibinfo{person}{Jianliang Xiao},
  \bibinfo{person}{Robin Deterding}, \bibinfo{person}{Thang Dinh}, {and}
  \bibinfo{person}{Tam Vu}.} \bibinfo{year}{2021}\natexlab{}.
\newblock \showarticletitle{EBP: An Ear-Worn Device for Frequent and
  Comfortable Blood Pressure Monitoring}.
\newblock \bibinfo{journal}{{\em Commun. ACM\/}} (\bibinfo{year}{2021}).
\newblock


\bibitem[\protect\citeauthoryear{Chan, Ali, Najafi, Meehan, Mancl, Gallagher,
  Bly, and Gollakota}{Chan et~al\mbox{.}}{2022}]%
        {oae}
\bibfield{author}{\bibinfo{person}{Justin Chan}, \bibinfo{person}{Nada Ali},
  \bibinfo{person}{Ali Najafi}, \bibinfo{person}{Anna Meehan},
  \bibinfo{person}{Lisa Mancl}, \bibinfo{person}{Emily Gallagher},
  \bibinfo{person}{Randall Bly}, {and} \bibinfo{person}{Shyamnath Gollakota}.}
  \bibinfo{year}{2022}\natexlab{}.
\newblock \showarticletitle{An off-the-shelf otoacoustic-emission probe for
  hearing screening via a smartphone}.
\newblock \bibinfo{journal}{{\em Nature Biomedical Engineering\/}}
  \bibinfo{volume}{6} (\bibinfo{date}{10} \bibinfo{year}{2022}),
  \bibinfo{pages}{1--11}.
\newblock
\showDOI{%
\url{https://doi.org/10.1038/s41551-022-00947-6}}


\bibitem[\protect\citeauthoryear{Chan, Raju, Nandakumar, Bly, and
  Gollakota}{Chan et~al\mbox{.}}{2019}]%
        {infection}
\bibfield{author}{\bibinfo{person}{Justin Chan}, \bibinfo{person}{Sharat Raju},
  \bibinfo{person}{Rajalakshmi Nandakumar}, \bibinfo{person}{Randall Bly},
  {and} \bibinfo{person}{Shyamnath Gollakota}.}
  \bibinfo{year}{2019}\natexlab{}.
\newblock \showarticletitle{Detecting middle ear fluid using smartphones}.
\newblock \bibinfo{journal}{{\em Science Translational Medicine\/}}
  \bibinfo{volume}{11} (\bibinfo{date}{05} \bibinfo{year}{2019}),
  \bibinfo{pages}{eaav1102}.
\newblock
\showDOI{%
\url{https://doi.org/10.1126/scitranslmed.aav1102}}


\bibitem[\protect\citeauthoryear{Chatterjee, Kim, Jayaram, Gollakota,
  Kemelmacher, Patel, and Seitz}{Chatterjee et~al\mbox{.}}{2022}]%
        {chatterjee2022clearbuds}
\bibfield{author}{\bibinfo{person}{Ishan Chatterjee}, \bibinfo{person}{Maruchi
  Kim}, \bibinfo{person}{Vivek Jayaram}, \bibinfo{person}{Shyamnath Gollakota},
  \bibinfo{person}{Ira Kemelmacher}, \bibinfo{person}{Shwetak Patel}, {and}
  \bibinfo{person}{Steven~M Seitz}.} \bibinfo{year}{2022}\natexlab{}.
\newblock \showarticletitle{ClearBuds: wireless binaural earbuds for
  learning-based speech enhancement}. In \bibinfo{booktitle}{{\em ACM
  MobiSys}}.
\newblock


\bibitem[\protect\citeauthoryear{Delcroix, V{\'a}zquez, Ochiai, Kinoshita,
  Ohishi, and Araki}{Delcroix et~al\mbox{.}}{2022}]%
        {delcroix2022soundbeam}
\bibfield{author}{\bibinfo{person}{Marc Delcroix},
  \bibinfo{person}{Jorge~Bennasar V{\'a}zquez}, \bibinfo{person}{Tsubasa
  Ochiai}, \bibinfo{person}{Keisuke Kinoshita}, \bibinfo{person}{Yasunori
  Ohishi}, {and} \bibinfo{person}{Shoko Araki}.}
  \bibinfo{year}{2022}\natexlab{}.
\newblock \showarticletitle{SoundBeam: Target sound extraction conditioned on
  sound-class labels and enrollment clues for increased performance and
  continuous learning}. In \bibinfo{booktitle}{{\em arXiv}}.
\newblock


\bibitem[\protect\citeauthoryear{Doclo, Gannot, Moonen, Spriet, Haykin, and
  Liu}{Doclo et~al\mbox{.}}{2010}]%
        {doclo2010acoustic}
\bibfield{author}{\bibinfo{person}{Simon Doclo}, \bibinfo{person}{Sharon
  Gannot}, \bibinfo{person}{Marc Moonen}, \bibinfo{person}{Ann Spriet},
  \bibinfo{person}{Simon Haykin}, {and} \bibinfo{person}{KJ~Ray Liu}.}
  \bibinfo{year}{2010}\natexlab{}.
\newblock \showarticletitle{Acoustic beamforming for hearing aid applications}.
\newblock \bibinfo{journal}{{\em Handbook on array processing and sensor
  networks\/}} (\bibinfo{year}{2010}), \bibinfo{pages}{269--302}.
\newblock


\bibitem[\protect\citeauthoryear{Eskimez, Yoshioka, Wang, Wang, Chen, and
  Huang}{Eskimez et~al\mbox{.}}{2022}]%
        {eskimez2022personalized}
\bibfield{author}{\bibinfo{person}{Sefik~Emre Eskimez}, \bibinfo{person}{Takuya
  Yoshioka}, \bibinfo{person}{Huaming Wang}, \bibinfo{person}{Xiaofei Wang},
  \bibinfo{person}{Zhuo Chen}, {and} \bibinfo{person}{Xuedong Huang}.}
  \bibinfo{year}{2022}\natexlab{}.
\newblock \showarticletitle{Personalized speech enhancement: New models and
  comprehensive evaluation}. In \bibinfo{booktitle}{{\em IEEE ICASSP}}.
\newblock


\bibitem[\protect\citeauthoryear{Fonseca, Favory, Pons, Font, and
  Serra}{Fonseca et~al\mbox{.}}{2022}]%
        {fsd50k}
\bibfield{author}{\bibinfo{person}{Eduardo Fonseca}, \bibinfo{person}{Xavier
  Favory}, \bibinfo{person}{Jordi Pons}, \bibinfo{person}{Frederic Font}, {and}
  \bibinfo{person}{Xavier Serra}.} \bibinfo{year}{2022}\natexlab{}.
\newblock \bibinfo{title}{{FSD50K}: An Open Dataset of Human-Labeled Sound
  Events}.
\newblock   (\bibinfo{year}{2022}).
\newblock
\showeprint[arxiv]{cs.SD/2010.00475}


\bibitem[\protect\citeauthoryear{Gao and Grauman}{Gao and Grauman}{2019}]%
        {gao2019co}
\bibfield{author}{\bibinfo{person}{Ruohan Gao} {and} \bibinfo{person}{Kristen
  Grauman}.} \bibinfo{year}{2019}\natexlab{}.
\newblock \showarticletitle{Co-separating sounds of visual objects}. In
  \bibinfo{booktitle}{{\em IEEE /CVF ICCV}}.
\newblock


\bibitem[\protect\citeauthoryear{Gemmeke, Ellis, Freedman, Jansen, Lawrence,
  Moore, Plakal, and Ritter}{Gemmeke et~al\mbox{.}}{2017}]%
        {audioset}
\bibfield{author}{\bibinfo{person}{Jort~F. Gemmeke}, \bibinfo{person}{Daniel
  P.~W. Ellis}, \bibinfo{person}{Dylan Freedman}, \bibinfo{person}{Aren
  Jansen}, \bibinfo{person}{Wade Lawrence}, \bibinfo{person}{R.~Channing
  Moore}, \bibinfo{person}{Manoj Plakal}, {and} \bibinfo{person}{Marvin
  Ritter}.} \bibinfo{year}{2017}\natexlab{}.
\newblock \showarticletitle{Audio Set: An ontology and human-labeled dataset
  for audio events}. In \bibinfo{booktitle}{{\em IEEE ICASSP}}.
\newblock


\bibitem[\protect\citeauthoryear{Gfeller, Roblek, and Tagliasacchi}{Gfeller
  et~al\mbox{.}}{2021}]%
        {gfeller2021one}
\bibfield{author}{\bibinfo{person}{Beat Gfeller}, \bibinfo{person}{Dominik
  Roblek}, {and} \bibinfo{person}{Marco Tagliasacchi}.}
  \bibinfo{year}{2021}\natexlab{}.
\newblock \showarticletitle{One-shot conditional audio filtering of arbitrary
  sounds}. In \bibinfo{booktitle}{{\em ICASSP}}. IEEE.
\newblock


\bibitem[\protect\citeauthoryear{Giri, Venkataramani, Valin, Isik, and
  Krishnaswamy}{Giri et~al\mbox{.}}{2021}]%
        {giri2021personalized}
\bibfield{author}{\bibinfo{person}{Ritwik Giri}, \bibinfo{person}{Shrikant
  Venkataramani}, \bibinfo{person}{Jean-Marc Valin}, \bibinfo{person}{Umut
  Isik}, {and} \bibinfo{person}{Arvindh Krishnaswamy}.}
  \bibinfo{year}{2021}\natexlab{}.
\newblock \showarticletitle{Personalized percepnet: Real-time, low-complexity
  target voice separation and enhancement}. In \bibinfo{booktitle}{{\em
  arXiv}}.
\newblock


\bibitem[\protect\citeauthoryear{Gu, Wu, Zhang, Chen, Xu, Yu, Su, Zou, and
  Yu}{Gu et~al\mbox{.}}{2019}]%
        {gu2019endtoend}
\bibfield{author}{\bibinfo{person}{Rongzhi Gu}, \bibinfo{person}{Jian Wu},
  \bibinfo{person}{Shi-Xiong Zhang}, \bibinfo{person}{Lianwu Chen},
  \bibinfo{person}{Yong Xu}, \bibinfo{person}{Meng Yu}, \bibinfo{person}{Dan
  Su}, \bibinfo{person}{Yuexian Zou}, {and} \bibinfo{person}{Dong Yu}.}
  \bibinfo{year}{2019}\natexlab{}.
\newblock \showarticletitle{End-to-End Multi-Channel Speech Separation}. In
  \bibinfo{booktitle}{{\em arXiv}}.
\newblock
\showeprint{cs.SD/1905.06286}


\bibitem[\protect\citeauthoryear{Gupta, Ranjan, He, Gan, and Peksi}{Gupta
  et~al\mbox{.}}{2020}]%
        {gupta2020acoustic}
\bibfield{author}{\bibinfo{person}{Rishabh Gupta}, \bibinfo{person}{Rishabh
  Ranjan}, \bibinfo{person}{Jianjun He}, \bibinfo{person}{Woon-Seng Gan}, {and}
  \bibinfo{person}{Santi Peksi}.} \bibinfo{year}{2020}\natexlab{}.
\newblock \showarticletitle{Acoustic transparency in hearables for augmented
  reality audio: Hear-through techniques review and challenges}. In
  \bibinfo{booktitle}{{\em Audio Engineering Society Conference on Audio for
  Virtual and Augmented Reality}}.
\newblock


\bibitem[\protect\citeauthoryear{Han, Luo, and Mesgarani}{Han
  et~al\mbox{.}}{2020}]%
        {han2020realtime}
\bibfield{author}{\bibinfo{person}{Cong Han}, \bibinfo{person}{Yi Luo}, {and}
  \bibinfo{person}{Nima Mesgarani}.} \bibinfo{year}{2020}\natexlab{}.
\newblock \showarticletitle{Real-time binaural speech separation with preserved
  spatial cues}. In \bibinfo{booktitle}{{\em arXiv}}.
\newblock
\showeprint{eess.AS/2002.06637}


\bibitem[\protect\citeauthoryear{IoSR-Surrey}{IoSR-Surrey}{2016}]%
        {rrbrir}
\bibfield{author}{\bibinfo{person}{IoSR-Surrey}.}
  \bibinfo{year}{2016}\natexlab{}.
\newblock \bibinfo{title}{IoSR-surrey/realroombrirs: Binaural impulse responses
  captured in real rooms.}
\newblock
  \bibinfo{howpublished}{\url{https://github.com/IoSR-Surrey/RealRoomBRIRs}}.
  (\bibinfo{year}{2016}).
\newblock


\bibitem[\protect\citeauthoryear{IoSR-Surrey}{IoSR-Surrey}{2023}]%
        {CATT_RIR}
\bibfield{author}{\bibinfo{person}{IoSR-Surrey}.}
  \bibinfo{year}{2023}\natexlab{}.
\newblock \bibinfo{title}{Simulated Room Impulse Responses.}
\newblock \bibinfo{howpublished}{\url{https://iosr.uk/software/index.php}}.
  (\bibinfo{year}{2023}).
\newblock


\bibitem[\protect\citeauthoryear{Jain, Mack, Amrous, Wright, Goodman,
  Findlater, and Froehlich}{Jain et~al\mbox{.}}{2020a}]%
        {dhruv2}
\bibfield{author}{\bibinfo{person}{Dhruv Jain}, \bibinfo{person}{Kelly Mack},
  \bibinfo{person}{Akli Amrous}, \bibinfo{person}{Matt Wright},
  \bibinfo{person}{Steven Goodman}, \bibinfo{person}{Leah Findlater}, {and}
  \bibinfo{person}{Jon~E. Froehlich}.} \bibinfo{year}{2020}\natexlab{a}.
\newblock \showarticletitle{HomeSound: An Iterative Field Deployment of an
  In-Home Sound Awareness System for Deaf or Hard of Hearing Users}. In
  \bibinfo{booktitle}{{\em ACM CHI}}.
\newblock


\bibitem[\protect\citeauthoryear{Jain, Ngo, Patel, Goodman, Findlater, and
  Froehlich}{Jain et~al\mbox{.}}{2020b}]%
        {dhruv}
\bibfield{author}{\bibinfo{person}{Dhruv Jain}, \bibinfo{person}{Hung Ngo},
  \bibinfo{person}{Pratyush Patel}, \bibinfo{person}{Steven Goodman},
  \bibinfo{person}{Leah Findlater}, {and} \bibinfo{person}{Jon Froehlich}.}
  \bibinfo{year}{2020}\natexlab{b}.
\newblock \showarticletitle{SoundWatch: Exploring Smartwatch-Based Deep
  Learning Approaches to Support Sound Awareness for Deaf and Hard of Hearing
  Users}. In \bibinfo{booktitle}{{\em ACM SIGACCESS ASSETS}}.
\newblock


\bibitem[\protect\citeauthoryear{Jansson, Humphrey, Montecchio, Bittner, Kumar,
  and Weyde}{Jansson et~al\mbox{.}}{2017}]%
        {Jansson2017SingingVS}
\bibfield{author}{\bibinfo{person}{Andreas Jansson}, \bibinfo{person}{Eric~J.
  Humphrey}, \bibinfo{person}{Nicola Montecchio}, \bibinfo{person}{Rachel~M.
  Bittner}, \bibinfo{person}{Aparna Kumar}, {and} \bibinfo{person}{Tillman
  Weyde}.} \bibinfo{year}{2017}\natexlab{}.
\newblock \showarticletitle{Singing Voice Separation with Deep U-Net
  Convolutional Networks}. In \bibinfo{booktitle}{{\em ISMIR}}.
\newblock


\bibitem[\protect\citeauthoryear{Jin, Schoof, and Schepker}{Jin
  et~al\mbox{.}}{2022}]%
        {transparency2}
\bibfield{author}{\bibinfo{person}{Wenyu Jin}, \bibinfo{person}{Tim Schoof},
  {and} \bibinfo{person}{Henning Schepker}.} \bibinfo{year}{2022}\natexlab{}.
\newblock \showarticletitle{Individualized Hear-Through For Acoustic
  Transparency Using {PCA}-Based Sound Pressure Estimation At The Eardrum}. In
  \bibinfo{booktitle}{{\em ICASSP}}.
\newblock


\bibitem[\protect\citeauthoryear{Jorgewich-Cohen, Townsend, Padovese, Klein,
  Praschag, Ferrara, Ettmar, Menezes, Varani, Serano, and
  Sánchez-Villagra}{Jorgewich-Cohen et~al\mbox{.}}{2022}]%
        {hearingnature}
\bibfield{author}{\bibinfo{person}{Gabriel Jorgewich-Cohen},
  \bibinfo{person}{Simon Townsend}, \bibinfo{person}{Linilson Padovese},
  \bibinfo{person}{Nicole Klein}, \bibinfo{person}{Peter Praschag},
  \bibinfo{person}{Camila Ferrara}, \bibinfo{person}{Stephan Ettmar},
  \bibinfo{person}{Sabrina Menezes}, \bibinfo{person}{Arthur Varani},
  \bibinfo{person}{Jaren Serano}, {and} \bibinfo{person}{Marcelo
  Sánchez-Villagra}.} \bibinfo{year}{2022}\natexlab{}.
\newblock \showarticletitle{Common evolutionary origin of acoustic
  communication in choanate vertebrates}.
\newblock \bibinfo{journal}{{\em Nature Communications\/}}
  \bibinfo{volume}{13} (\bibinfo{date}{10} \bibinfo{year}{2022}).
\newblock
\showDOI{%
\url{https://doi.org/10.1038/s41467-022-33741-8}}


\bibitem[\protect\citeauthoryear{Kilgour, Gfeller, Huang, Jansen, Wisdom, and
  Tagliasacchi}{Kilgour et~al\mbox{.}}{2022}]%
        {kilgour2022text}
\bibfield{author}{\bibinfo{person}{Kevin Kilgour}, \bibinfo{person}{Beat
  Gfeller}, \bibinfo{person}{Qingqing Huang}, \bibinfo{person}{Aren Jansen},
  \bibinfo{person}{Scott Wisdom}, {and} \bibinfo{person}{Marco Tagliasacchi}.}
  \bibinfo{year}{2022}\natexlab{}.
\newblock \showarticletitle{Text-Driven Separation of Arbitrary Sounds}. In
  \bibinfo{booktitle}{{\em arXiv}}.
\newblock


\bibitem[\protect\citeauthoryear{Laput, Ahuja, Goel, and Harrison}{Laput
  et~al\mbox{.}}{2018}]%
        {ubicoustics}
\bibfield{author}{\bibinfo{person}{Gierad Laput}, \bibinfo{person}{Karan
  Ahuja}, \bibinfo{person}{Mayank Goel}, {and} \bibinfo{person}{Chris
  Harrison}.} \bibinfo{year}{2018}\natexlab{}.
\newblock \showarticletitle{Ubicoustics: Plug-and-Play Acoustic Activity
  Recognition}. In \bibinfo{booktitle}{{\em ACM UIST}}.
\newblock


\bibitem[\protect\citeauthoryear{Liu, Liu, Kong, Mei, Zhao, Huang, Plumbley,
  and Wang}{Liu et~al\mbox{.}}{2022}]%
        {liu2022separate}
\bibfield{author}{\bibinfo{person}{Xubo Liu}, \bibinfo{person}{Haohe Liu},
  \bibinfo{person}{Qiuqiang Kong}, \bibinfo{person}{Xinhao Mei},
  \bibinfo{person}{Jinzheng Zhao}, \bibinfo{person}{Qiushi Huang},
  \bibinfo{person}{Mark~D Plumbley}, {and} \bibinfo{person}{Wenwu Wang}.}
  \bibinfo{year}{2022}\natexlab{}.
\newblock \showarticletitle{Separate What You Describe: Language-Queried Audio
  Source Separation}. In \bibinfo{booktitle}{{\em arXiv}}.
\newblock


\bibitem[\protect\citeauthoryear{Lu, Pan, Lane, Choudhury, and Campbell}{Lu
  et~al\mbox{.}}{2009}]%
        {soundsense}
\bibfield{author}{\bibinfo{person}{Hong Lu}, \bibinfo{person}{Wei Pan},
  \bibinfo{person}{Nicholas~D. Lane}, \bibinfo{person}{Tanzeem Choudhury},
  {and} \bibinfo{person}{Andrew~T. Campbell}.} \bibinfo{year}{2009}\natexlab{}.
\newblock \showarticletitle{SoundSense: Scalable Sound Sensing for
  People-Centric Applications on Mobile Phones}. In \bibinfo{booktitle}{{\em
  ACM MobiSys}}.
\newblock


\bibitem[\protect\citeauthoryear{Luo, Wang, Cheng, Xiao, Zhang, and Xiao}{Luo
  et~al\mbox{.}}{2022}]%
        {luo2022tiny}
\bibfield{author}{\bibinfo{person}{Jian Luo}, \bibinfo{person}{Jianzong Wang},
  \bibinfo{person}{Ning Cheng}, \bibinfo{person}{Edward Xiao},
  \bibinfo{person}{Xulong Zhang}, {and} \bibinfo{person}{Jing Xiao}.}
  \bibinfo{year}{2022}\natexlab{}.
\newblock \showarticletitle{Tiny-Sepformer: A Tiny Time-Domain Transformer
  Network for Speech Separation}. In \bibinfo{booktitle}{{\em arXiv}}.
\newblock


\bibitem[\protect\citeauthoryear{Luo and Mesgarani}{Luo and Mesgarani}{2019}]%
        {luo2019conv}
\bibfield{author}{\bibinfo{person}{Yi Luo} {and} \bibinfo{person}{Nima
  Mesgarani}.} \bibinfo{year}{2019}\natexlab{}.
\newblock \showarticletitle{Conv-tasnet: Surpassing ideal time--frequency
  magnitude masking for speech separation}.
\newblock \bibinfo{journal}{{\em IEEE/ACM transactions on audio, speech, and
  language processing\/}} (\bibinfo{year}{2019}).
\newblock


\bibitem[\protect\citeauthoryear{Ma, Ferlini, and Mascolo}{Ma
  et~al\mbox{.}}{2021}]%
        {mobisys21}
\bibfield{author}{\bibinfo{person}{Dong Ma}, \bibinfo{person}{Andrea Ferlini},
  {and} \bibinfo{person}{Cecilia Mascolo}.} \bibinfo{year}{2021}\natexlab{}.
\newblock \showarticletitle{{OES}ense: Employing Occlusion Effect for in-Ear
  Human Sensing}. In \bibinfo{booktitle}{{\em MobiSys}}.
\newblock


\bibitem[\protect\citeauthoryear{May, van~de Par, and Kohlrausch}{May
  et~al\mbox{.}}{2011}]%
        {itd_matlab}
\bibfield{author}{\bibinfo{person}{Tobias May}, \bibinfo{person}{Steven van~de
  Par}, {and} \bibinfo{person}{Armin Kohlrausch}.}
  \bibinfo{year}{2011}\natexlab{}.
\newblock \showarticletitle{A Probabilistic Model for Robust Localization Based
  on a Binaural Auditory Front-End}.
\newblock \bibinfo{journal}{{\em IEEE Transactions on Audio, Speech, and
  Language Processing\/}} \bibinfo{volume}{19}, \bibinfo{number}{1}
  (\bibinfo{year}{2011}), \bibinfo{pages}{1--13}.
\newblock
\showDOI{%
\url{https://doi.org/10.1109/TASL.2010.2042128}}


\bibitem[\protect\citeauthoryear{McDonnell, Moon, Jiang, Goodman, Kushalnaga,
  Froehlich, and Findlater}{McDonnell et~al\mbox{.}}{2023}]%
        {jon}
\bibfield{author}{\bibinfo{person}{Emma McDonnell}, \bibinfo{person}{Soo~Hyun
  Moon}, \bibinfo{person}{Lucy Jiang}, \bibinfo{person}{Steven Goodman},
  \bibinfo{person}{Raja Kushalnaga}, \bibinfo{person}{Jon Froehlich}, {and}
  \bibinfo{person}{Leah Findlater}.} \bibinfo{year}{2023}\natexlab{}.
\newblock \showarticletitle{“Easier or Harder, Depending on Who the Hearing
  Person Is”: Codesigning Videoconferencing Tools for Small Groups with Mixed
  Hearing Status}. In \bibinfo{booktitle}{{\em ACM CHI}}.
\newblock


\bibitem[\protect\citeauthoryear{Mesaros, Heittola, and Virtanen}{Mesaros
  et~al\mbox{.}}{2018}]%
        {Mesaros2018_DCASE}
\bibfield{author}{\bibinfo{person}{Annamaria Mesaros}, \bibinfo{person}{Toni
  Heittola}, {and} \bibinfo{person}{Tuomas Virtanen}.}
  \bibinfo{year}{2018}\natexlab{}.
\newblock \showarticletitle{A multi-device dataset for urban acoustic scene
  classification}. In \bibinfo{booktitle}{{\em DCASE}}.
\newblock
\showURL{%
\url{https://arxiv.org/abs/1807.09840}}


\bibitem[\protect\citeauthoryear{Mollyn, Ahuja, Verma, Harrison, and
  Goel}{Mollyn et~al\mbox{.}}{2022}]%
        {samosa}
\bibfield{author}{\bibinfo{person}{Vimal Mollyn}, \bibinfo{person}{Karan
  Ahuja}, \bibinfo{person}{Dhruv Verma}, \bibinfo{person}{Chris Harrison},
  {and} \bibinfo{person}{Mayank Goel}.} \bibinfo{year}{2022}\natexlab{}.
\newblock \showarticletitle{SAMoSA: Sensing Activities with Motion and
  Subsampled Audio}.
\newblock \bibinfo{journal}{{\em IMWUT\/}} (\bibinfo{year}{2022}).
\newblock


\bibitem[\protect\citeauthoryear{Nicolas}{Nicolas}{2020}]%
        {disco}
\bibfield{author}{\bibinfo{person}{Furnon Nicolas}.}
  \bibinfo{year}{2020}\natexlab{}.
\newblock \bibinfo{title}{Noise files for the DISCO dataset}.
\newblock   (\bibinfo{year}{2020}).
\newblock
\newblock
\shownote{\url{https://github.com/nfurnon/disco}.}


\bibitem[\protect\citeauthoryear{{Ochiai}, {Delcroix}, {Koizumi}, {Ito},
  {Kinoshita}, and {Araki}}{{Ochiai} et~al\mbox{.}}{2020}]%
        {2020arXiv200605712O}
\bibfield{author}{\bibinfo{person}{Tsubasa {Ochiai}}, \bibinfo{person}{Marc
  {Delcroix}}, \bibinfo{person}{Yuma {Koizumi}}, \bibinfo{person}{Hiroaki
  {Ito}}, \bibinfo{person}{Keisuke {Kinoshita}}, {and} \bibinfo{person}{Shoko
  {Araki}}.} \bibinfo{year}{2020}\natexlab{}.
\newblock \showarticletitle{{Listen to What You Want: Neural Network-based
  Universal Sound Selector}}, In \bibinfo{booktitle}{arXiv}.
\newblock \bibinfo{journal}{{\em arXiv e-prints\/}}.
\newblock
\showeprint{eess.AS/2006.05712}


\bibitem[\protect\citeauthoryear{Okamoto, Horiguchi, Yamamoto, Imoto, and
  Kawaguchi}{Okamoto et~al\mbox{.}}{2022}]%
        {okamoto2022environmental}
\bibfield{author}{\bibinfo{person}{Yuki Okamoto}, \bibinfo{person}{Shota
  Horiguchi}, \bibinfo{person}{Masaaki Yamamoto}, \bibinfo{person}{Keisuke
  Imoto}, {and} \bibinfo{person}{Yohei Kawaguchi}.}
  \bibinfo{year}{2022}\natexlab{}.
\newblock \showarticletitle{Environmental Sound Extraction Using Onomatopoeic
  Words}. In \bibinfo{booktitle}{{\em IEEE ICASSP}}.
\newblock


\bibitem[\protect\citeauthoryear{Oord, Dieleman, Zen, Simonyan, Vinyals,
  Graves, Kalchbrenner, Senior, and Kavukcuoglu}{Oord et~al\mbox{.}}{2016}]%
        {https://doi.org/10.48550/arxiv.1609.03499}
\bibfield{author}{\bibinfo{person}{Aaron van~den Oord}, \bibinfo{person}{Sander
  Dieleman}, \bibinfo{person}{Heiga Zen}, \bibinfo{person}{Karen Simonyan},
  \bibinfo{person}{Oriol Vinyals}, \bibinfo{person}{Alex Graves},
  \bibinfo{person}{Nal Kalchbrenner}, \bibinfo{person}{Andrew Senior}, {and}
  \bibinfo{person}{Koray Kavukcuoglu}.} \bibinfo{year}{2016}\natexlab{}.
\newblock \showarticletitle{WaveNet: A Generative Model for Raw Audio}. In
  \bibinfo{booktitle}{{\em arXiv}}.
\newblock
\showDOI{%
\url{https://doi.org/10.48550/ARXIV.1609.03499}}


\bibitem[\protect\citeauthoryear{Paine, Khorrami, Chang, Zhang, Ramachandran,
  Hasegawa-Johnson, and Huang}{Paine et~al\mbox{.}}{2016}]%
        {paine2016fast}
\bibfield{author}{\bibinfo{person}{Tom~Le Paine}, \bibinfo{person}{Pooya
  Khorrami}, \bibinfo{person}{Shiyu Chang}, \bibinfo{person}{Yang Zhang},
  \bibinfo{person}{Prajit Ramachandran}, \bibinfo{person}{Mark~A.
  Hasegawa-Johnson}, {and} \bibinfo{person}{Thomas~S. Huang}.}
  \bibinfo{year}{2016}\natexlab{}.
\newblock \showarticletitle{Fast Wavenet Generation Algorithm}. In
  \bibinfo{booktitle}{{\em arXiv}}.
\newblock
\showeprint{cs.SD/1611.09482}


\bibitem[\protect\citeauthoryear{Pavel, Reyes, and Bigham}{Pavel
  et~al\mbox{.}}{2020}]%
        {editing2}
\bibfield{author}{\bibinfo{person}{Amy Pavel}, \bibinfo{person}{Gabriel Reyes},
  {and} \bibinfo{person}{Jeffrey~P. Bigham}.} \bibinfo{year}{2020}\natexlab{}.
\newblock \showarticletitle{Rescribe: Authoring and Automatically Editing Audio
  Descriptions}. In \bibinfo{booktitle}{{\em ACM UIST}}.
\newblock


\bibitem[\protect\citeauthoryear{Peterson}{Peterson}{2021}]%
        {airpodssales}
\bibfield{author}{\bibinfo{person}{Mike Peterson}.}
  \bibinfo{year}{2021}\natexlab{}.
\newblock \bibinfo{title}{Apple {A}ir{P}ods, {B}eats dominated audio wearable
  market in 2020}.
\newblock
  \bibinfo{howpublished}{\url{https://appleinsider.com/articles/21/03/30/apple-airpods-beats-dominated-audio-wearable-market-in-2020}}.
    (\bibinfo{year}{2021}).
\newblock


\bibitem[\protect\citeauthoryear{Piczak}{Piczak}{2015}]%
        {esc50}
\bibfield{author}{\bibinfo{person}{Karol~J. Piczak}.}
  \bibinfo{year}{2015}\natexlab{}.
\newblock \showarticletitle{ESC: Dataset for Environmental Sound
  Classification}. In \bibinfo{booktitle}{{\em ACM Multimedia}}.
\newblock


\bibitem[\protect\citeauthoryear{Prakash, Yang, Wei, Hassanieh, and
  Choudhury}{Prakash et~al\mbox{.}}{2020}]%
        {teeth}
\bibfield{author}{\bibinfo{person}{Jay Prakash}, \bibinfo{person}{Zhijian
  Yang}, \bibinfo{person}{Yu-Lin Wei}, \bibinfo{person}{Haitham Hassanieh},
  {and} \bibinfo{person}{Romit~Roy Choudhury}.}
  \bibinfo{year}{2020}\natexlab{}.
\newblock \showarticletitle{Ear{S}ense: Earphones as a Teeth Activity Sensor}.
  In \bibinfo{booktitle}{{\em MobiCom}}.
\newblock


\bibitem[\protect\citeauthoryear{Rafii, Liutkus, Stöter, Mimilakis, and
  Bittner}{Rafii et~al\mbox{.}}{2017}]%
        {musdb18}
\bibfield{author}{\bibinfo{person}{Zafar Rafii}, \bibinfo{person}{Antoine
  Liutkus}, \bibinfo{person}{Fabian-Robert Stöter},
  \bibinfo{person}{Stylianos~Ioannis Mimilakis}, {and} \bibinfo{person}{Rachel
  Bittner}.} \bibinfo{year}{2017}\natexlab{}.
\newblock \bibinfo{title}{MUSDB18 - a corpus for music separation}.
\newblock   (\bibinfo{year}{2017}).
\newblock


\bibitem[\protect\citeauthoryear{Roux, Wisdom, Erdogan, and Hershey}{Roux
  et~al\mbox{.}}{2018}]%
        {https://doi.org/10.48550/arxiv.1811.02508}
\bibfield{author}{\bibinfo{person}{Jonathan~Le Roux}, \bibinfo{person}{Scott
  Wisdom}, \bibinfo{person}{Hakan Erdogan}, {and} \bibinfo{person}{John~R.
  Hershey}.} \bibinfo{year}{2018}\natexlab{}.
\newblock \showarticletitle{{SDR} - half-baked or well done?}. In
  \bibinfo{booktitle}{{\em arXiv}}.
\newblock


\bibitem[\protect\citeauthoryear{Rubin, Berthouzoz, Mysore, Li, and
  Agrawala}{Rubin et~al\mbox{.}}{2013}]%
        {editing1}
\bibfield{author}{\bibinfo{person}{Steve Rubin}, \bibinfo{person}{Floraine
  Berthouzoz}, \bibinfo{person}{Gautham~J. Mysore}, \bibinfo{person}{Wilmot
  Li}, {and} \bibinfo{person}{Maneesh Agrawala}.}
  \bibinfo{year}{2013}\natexlab{}.
\newblock \showarticletitle{Content-Based Tools for Editing Audio Stories}. In
  \bibinfo{booktitle}{{\em ACM UIST}}.
\newblock


\bibitem[\protect\citeauthoryear{Salamon, MacConnell, Cartwright, Li, and
  Bello}{Salamon et~al\mbox{.}}{2017}]%
        {8170052}
\bibfield{author}{\bibinfo{person}{Justin Salamon}, \bibinfo{person}{Duncan
  MacConnell}, \bibinfo{person}{Mark Cartwright}, \bibinfo{person}{Peter Li},
  {and} \bibinfo{person}{Juan~Pablo Bello}.} \bibinfo{year}{2017}\natexlab{}.
\newblock \showarticletitle{Scaper: A library for soundscape synthesis and
  augmentation}. In \bibinfo{booktitle}{{\em WASPAA}}.
\newblock
\showDOI{%
\url{https://doi.org/10.1109/WASPAA.2017.8170052}}


\bibitem[\protect\citeauthoryear{Satongar, Lam, Pike, et~al\mbox{.}}{Satongar
  et~al\mbox{.}}{2014}]%
        {sbsbrir}
\bibfield{author}{\bibinfo{person}{Darius Satongar}, \bibinfo{person}{Yiu~W
  Lam}, \bibinfo{person}{Chris Pike}, {et~al\mbox{.}}}
  \bibinfo{year}{2014}\natexlab{}.
\newblock \showarticletitle{The {S}alford {BBC} Spatially-sampled Binaural Room
  Impulse Response dataset}.
\newblock  (\bibinfo{year}{2014}).
\newblock


\bibitem[\protect\citeauthoryear{Shen, Roy, Guan, Hassanieh, and
  Choudhury}{Shen et~al\mbox{.}}{2018}]%
        {cancel}
\bibfield{author}{\bibinfo{person}{Sheng Shen}, \bibinfo{person}{Nirupam Roy},
  \bibinfo{person}{Junfeng Guan}, \bibinfo{person}{Haitham Hassanieh}, {and}
  \bibinfo{person}{Romit~Roy Choudhury}.} \bibinfo{year}{2018}\natexlab{}.
\newblock \showarticletitle{{MUTE}: Bringing {I}o{T} to Noise Cancellation}. In
  \bibinfo{booktitle}{{\em ACM SIGCOMM}}.
\newblock
\showISBNx{9781450355674}
\showDOI{%
\url{https://doi.org/10.1145/3230543.3230550}}


\bibitem[\protect\citeauthoryear{Stone and Moore}{Stone and Moore}{1999}]%
        {stone1999tolerable}
\bibfield{author}{\bibinfo{person}{Michael~A Stone} {and}
  \bibinfo{person}{Brian~CJ Moore}.} \bibinfo{year}{1999}\natexlab{}.
\newblock \showarticletitle{Tolerable hearing aid delays. I. Estimation of
  limits imposed by the auditory path alone using simulated hearing losses}.
\newblock \bibinfo{journal}{{\em Ear and Hearing\/}} \bibinfo{volume}{20},
  \bibinfo{number}{3} (\bibinfo{year}{1999}), \bibinfo{pages}{182--192}.
\newblock


\bibitem[\protect\citeauthoryear{Subakan, Ravanelli, Cornell, Bronzi, and
  Zhong}{Subakan et~al\mbox{.}}{2021}]%
        {subakan2021attention}
\bibfield{author}{\bibinfo{person}{Cem Subakan}, \bibinfo{person}{Mirco
  Ravanelli}, \bibinfo{person}{Samuele Cornell}, \bibinfo{person}{Mirko
  Bronzi}, {and} \bibinfo{person}{Jianyuan Zhong}.}
  \bibinfo{year}{2021}\natexlab{}.
\newblock \showarticletitle{Attention is all you need in speech separation}. In
  \bibinfo{booktitle}{{\em IEEE ICASSP}}.
\newblock


\bibitem[\protect\citeauthoryear{Subakan, Ravanelli, Cornell, Lepoutre, and
  Grondin}{Subakan et~al\mbox{.}}{2022}]%
        {subakan2022resource}
\bibfield{author}{\bibinfo{person}{Cem Subakan}, \bibinfo{person}{Mirco
  Ravanelli}, \bibinfo{person}{Samuele Cornell},
  \bibinfo{person}{Fr{\'e}d{\'e}ric Lepoutre}, {and}
  \bibinfo{person}{Fran{\c{c}}ois Grondin}.} \bibinfo{year}{2022}\natexlab{}.
\newblock \showarticletitle{Resource-Efficient Separation Transformer}. In
  \bibinfo{booktitle}{{\em arXiv}}.
\newblock


\bibitem[\protect\citeauthoryear{Tambe, Yang, Ko, Chai, Hooper, Donato,
  Whatmough, Rush, Brooks, and Wei}{Tambe et~al\mbox{.}}{2022}]%
        {silicon}
\bibfield{author}{\bibinfo{person}{Thierry Tambe}, \bibinfo{person}{En-Yu
  Yang}, \bibinfo{person}{Glenn Ko}, \bibinfo{person}{Yuji Chai},
  \bibinfo{person}{Coleman Hooper}, \bibinfo{person}{Marco Donato},
  \bibinfo{person}{Paul Whatmough}, \bibinfo{person}{Alexander Rush},
  \bibinfo{person}{David Brooks}, {and} \bibinfo{person}{Gu-Yeon Wei}.}
  \bibinfo{year}{2022}\natexlab{}.
\newblock \showarticletitle{A 16-nm {S}o{C} for Noise-Robust Speech and {NLP}
  Edge {AI} Inference With Bayesian Sound Source Separation and Attention-Based
  {DNN}s}.
\newblock \bibinfo{journal}{{\em IEEE Journal of Solid-State Circuits\/}}
  (\bibinfo{year}{2022}).
\newblock
\showDOI{%
\url{https://doi.org/10.1109/JSSC.2022.3179303}}


\bibitem[\protect\citeauthoryear{Tonami, Imoto, Nagase, Okamoto, Fukumori, and
  Yamashita}{Tonami et~al\mbox{.}}{2022}]%
        {tonami2022sound}
\bibfield{author}{\bibinfo{person}{Noriyuki Tonami}, \bibinfo{person}{Keisuke
  Imoto}, \bibinfo{person}{Ryotaro Nagase}, \bibinfo{person}{Yuki Okamoto},
  \bibinfo{person}{Takahiro Fukumori}, {and} \bibinfo{person}{Yoichi
  Yamashita}.} \bibinfo{year}{2022}\natexlab{}.
\newblock \showarticletitle{Sound Event Detection Guided by Semantic Contexts
  of Scenes}. In \bibinfo{booktitle}{{\em arXiv}}.
\newblock
\showeprint{cs.SD/2110.03243}


\bibitem[\protect\citeauthoryear{Valimaki, Franck, Ramo, Gamper, and
  Savioja}{Valimaki et~al\mbox{.}}{2015}]%
        {ieee}
\bibfield{author}{\bibinfo{person}{Vesa Valimaki}, \bibinfo{person}{Andreas
  Franck}, \bibinfo{person}{Jussi Ramo}, \bibinfo{person}{Hannes Gamper}, {and}
  \bibinfo{person}{Lauri Savioja}.} \bibinfo{year}{2015}\natexlab{}.
\newblock \showarticletitle{Assisted Listening Using a Headset: Enhancing audio
  perception in real, augmented, and virtual environments}.
\newblock \bibinfo{journal}{{\em IEEE Signal Processing Magazine\/}}
  \bibinfo{volume}{32}, \bibinfo{number}{2} (\bibinfo{year}{2015}),
  \bibinfo{pages}{92--99}.
\newblock
\showDOI{%
\url{https://doi.org/10.1109/MSP.2014.2369191}}


\bibitem[\protect\citeauthoryear{Vaswani, Shazeer, Parmar, Uszkoreit, Jones,
  Gomez, Kaiser, and Polosukhin}{Vaswani et~al\mbox{.}}{2017}]%
        {https://doi.org/10.48550/arxiv.1706.03762}
\bibfield{author}{\bibinfo{person}{Ashish Vaswani}, \bibinfo{person}{Noam
  Shazeer}, \bibinfo{person}{Niki Parmar}, \bibinfo{person}{Jakob Uszkoreit},
  \bibinfo{person}{Llion Jones}, \bibinfo{person}{Aidan~N. Gomez},
  \bibinfo{person}{Lukasz Kaiser}, {and} \bibinfo{person}{Illia Polosukhin}.}
  \bibinfo{year}{2017}\natexlab{}.
\newblock \showarticletitle{Attention Is All You Need}. In
  \bibinfo{booktitle}{{\em arXiv}}.
\newblock
\showDOI{%
\url{https://doi.org/10.48550/ARXIV.1706.03762}}


\bibitem[\protect\citeauthoryear{Veluri, Chan, Itani, Chen, Yoshioka, and
  Gollakota}{Veluri et~al\mbox{.}}{2023}]%
        {waveformer}
\bibfield{author}{\bibinfo{person}{Bandhav Veluri}, \bibinfo{person}{Justin
  Chan}, \bibinfo{person}{Malek Itani}, \bibinfo{person}{Tuochao Chen},
  \bibinfo{person}{Takuya Yoshioka}, {and} \bibinfo{person}{Shyamnath
  Gollakota}.} \bibinfo{year}{2023}\natexlab{}.
\newblock \showarticletitle{Real-Time Target Sound Extraction}. In
  \bibinfo{booktitle}{{\em IEEE ICASSP}}.
\newblock


\bibitem[\protect\citeauthoryear{Wang, Kim, Zhang, and Gollakota}{Wang
  et~al\mbox{.}}{2022b}]%
        {directional}
\bibfield{author}{\bibinfo{person}{Anran Wang}, \bibinfo{person}{Maruchi Kim},
  \bibinfo{person}{Hao Zhang}, {and} \bibinfo{person}{Shyamnath Gollakota}.}
  \bibinfo{year}{2022}\natexlab{b}.
\newblock \showarticletitle{Hybrid Neural Networks for On-Device Directional
  Hearing}.
\newblock \bibinfo{journal}{{\em AAAI\/}} (\bibinfo{year}{2022}).
\newblock
\showURL{%
\url{https://ojs.aaai.org/index.php/AAAI/article/view/21394}}


\bibitem[\protect\citeauthoryear{Wang, Ding, Chatterjee, Salemi~Parizi, Zhuang,
  Yan, Patel, and Shi}{Wang et~al\mbox{.}}{2022a}]%
        {chi22-ultrasonic}
\bibfield{author}{\bibinfo{person}{Yuntao Wang}, \bibinfo{person}{Jiexin Ding},
  \bibinfo{person}{Ishan Chatterjee}, \bibinfo{person}{Farshid Salemi~Parizi},
  \bibinfo{person}{Yuzhou Zhuang}, \bibinfo{person}{Yukang Yan},
  \bibinfo{person}{Shwetak Patel}, {and} \bibinfo{person}{Yuanchun Shi}.}
  \bibinfo{year}{2022}\natexlab{a}.
\newblock \showarticletitle{Face{O}ri: Tracking Head Position and Orientation
  Using Ultrasonic Ranging on Earphones}. In \bibinfo{booktitle}{{\em ACM
  CHI}}.
\newblock


\bibitem[\protect\citeauthoryear{Xu, Dai, and Lin}{Xu et~al\mbox{.}}{2019}]%
        {xu2019recursive}
\bibfield{author}{\bibinfo{person}{Xudong Xu}, \bibinfo{person}{Bo Dai}, {and}
  \bibinfo{person}{Dahua Lin}.} \bibinfo{year}{2019}\natexlab{}.
\newblock \showarticletitle{Recursive visual sound separation using minus-plus
  net}. In \bibinfo{booktitle}{{\em IEEE/CVF ICCV}}.
\newblock


\bibitem[\protect\citeauthoryear{Xu, Shi, Yi, Liu, Yan, Shi, Mariakakis,
  Mankoff, and Dey}{Xu et~al\mbox{.}}{2020}]%
        {earbuddy}
\bibfield{author}{\bibinfo{person}{Xuhai Xu}, \bibinfo{person}{Haitian Shi},
  \bibinfo{person}{Xin Yi}, \bibinfo{person}{WenJia Liu},
  \bibinfo{person}{Yukang Yan}, \bibinfo{person}{Yuanchun Shi},
  \bibinfo{person}{Alex Mariakakis}, \bibinfo{person}{Jennifer Mankoff}, {and}
  \bibinfo{person}{Anind~K. Dey}.} \bibinfo{year}{2020}\natexlab{}.
\newblock \showarticletitle{EarBuddy: Enabling On-Face Interaction via Wireless
  Earbuds}. In \bibinfo{booktitle}{{\em CHI}}.
\newblock
\showISBNx{9781450367080}
\showURL{%
\url{https://doi.org/10.1145/3313831.3376836}}


\bibitem[\protect\citeauthoryear{Yatani and Truong}{Yatani and Truong}{2012}]%
        {bodyscope}
\bibfield{author}{\bibinfo{person}{Koji Yatani} {and} \bibinfo{person}{Khai~N.
  Truong}.} \bibinfo{year}{2012}\natexlab{}.
\newblock \showarticletitle{BodyScope: A Wearable Acoustic Sensor for Activity
  Recognition}. In \bibinfo{booktitle}{{\em UbiComp}}.
\newblock


\end{thebibliography}

\end{document}